\documentclass[aps,preprint,showpacs,groupedaddress,floatfix]{revtex4}%
\usepackage{graphicx}
\usepackage{dcolumn}
\usepackage{bm}
\usepackage{amsmath}
\usepackage{amsfonts}
\usepackage{amssymb}%
\providecommand{\U}[1]{\protect\rule{.1in}{.1in}}

\newcommand{\be}{\begin{equation}}
\newcommand{\ee}{\end{equation}}
\newcommand{\bea}{\begin{eqnarray}}
\newcommand{\eea}{\end{eqnarray}}

\begin{document}
\title{Relativistic quasiparticle time blocking approximation. Dipole response of
open-shell nuclei}
\author{E. Litvinova}
\affiliation{Gesellschaft f\"{u}r Schwerionenforschung mbH, Planckstra{\ss }e 1, 64291
Darmstadt, Germany}
\affiliation{Institute of Physics and Power Engineering, 249033 Obninsk, Russia}
\author{P. Ring}
\affiliation{Physik-Department der Technischen Universit\"at M\"unchen, D-85748 Garching, Germany}
\affiliation{Departamento de F\'{\i}sica Te\'{o}rica, Universidad Aut\'{o}noma de Madrid,
E-28049 Madrid, Spain}
\author{V. Tselyaev}
\affiliation{Nuclear Physics Department, V. A. Fock Institute of Physics, St. Petersburg
State University, 198504, St. Petersburg, Russia}
\date{\today}

\begin{abstract}
The self-consistent Relativistic Quasiparticle Random Phase Approximation
(RQRPA) is extended by the quasiparticle-phonon coupling (QPC) model using the
Quasiparticle Time Blocking Approximation (QTBA). The method is formulated in
terms of the Bethe-Salpeter equation (BSE) in the two-quasiparticle space with
an energy-dependent two-quasiparticle residual interaction. This equation is
solved either in the basis of Dirac states forming the self-consistent
solution of the ground state or in the momentum representation. Pairing
correlations are treated within the Bardeen-Cooper-Schrieffer (BCS) model with
a monopole-monopole interaction. The same NL3 set of the coupling constants
generates the Dirac-Hartree-BCS single-quasiparticle spectrum, the static part
of the residual two-quasiparticle interaction and the quasiparticle-phonon
coupling amplitudes. A quantitative description of electric dipole excitations
in the chain of tin isotopes ($Z=50$) with the mass numbers $A=$ 100, 106,
114, 116, 120, and 130 and in the chain of isotones with ($N=50$) $^{88}$Sr,
$^{90}$Zr, $^{92}$Mo is performed within this framework.

The RQRPA extended by the coupling to collective vibrations generates spectra
with a multitude of $2q\otimes phonon$ (two quasiparticles plus phonon) states
providing a noticeable fragmentation of the giant dipole resonance as well as
of the soft dipole mode (pygmy resonance) in the nuclei under investigation.
The results obtained for the photo absorption cross sections and for the
integrated contributions of the low-lying strength to the calculated dipole
spectra agree very well with the available experimental data.

\end{abstract}

\maketitle


\section{Introduction}

Theoretical approaches based on Covariant Density Functional Theory (CDFT)
remain undoubtedly among the most successful microscopic descriptions of
nuclear structure. The CDFT approaches are derived from a Lorentz invariant
density functional which connects in a consistent way the spin and spatial
degrees of freedom in the nucleus. Therefore, it needs only a relatively small
number of parameters which are adjusted to reproduce a set of bulk properties
of spherical closed-shell nuclei~\cite{Rin.96,VALR.05} and it is valid over
the entire periodic table. Over the years, Relativistic Mean-Field (RMF)
models based on the CDFT have been successfully applied to describe ground
state properties of finite spherical and deformed nuclei over the entire
nuclear chart~\cite{GRT.90} from light nuclei~\cite{LVR.04a} to super-heavy
elements~\cite{LSRG.96,BBM.04}, from the neutron drip line where halo
phenomena are observed~\cite{MR.96}, to the proton drip line~\cite{LVR.04}
with nuclei unstable against the emission of protons ~\cite{LVR.99}. The
relativistic cranking approximation has been developed to calculate rotational
bands~\cite{KR.89,ARK.00}. For a description of nuclear excited states, the
Relativistic Random Phase Approximation (RRPA)~\cite{RMG.01} and the
quasiparticle RRPA (RQRPA)~\cite{PRN.03} have been formulated as the small
amplitude limit of the time-dependent RMF models. These models have provided a
very good description for the positions of giant resonances and a theoretical
interpretation of the low-lying dipole~\cite{PRN.03} and
quadrupole~\cite{Ans.05,AR.06} excitations. Proton-neutron versions of the
RRPA and the RQRPA have been developed and successfully applied to the
description of spin/isospin excitations as the Isobaric Analog Resonance (IAR)
or the Gamow-Teller Resonance (GTR)~\cite{PNV.04}.

Recently, several attempts have been made to extend the RMF and RRPA formalism
beyond the mean field approach, first of all, to solve the well known problem
of the RMF single-particle level density in the vicinity of the Fermi surface
which is too low because of the too small effective mass. The energy
dependence of the single-nucleon self-energy was emulated in a
phenomenological way~\cite{VNR.02} and microscopically by coupling the single
particle configurations to low-lying surface vibration~\cite{LR.06}. This
provided a considerable improvement for the description of the single-particle
spectra. An addition, the quadrupole motion has been studied within the
relativistic Generator Coordinate Method (GCM) ~\cite{NVR.06a,NVR.06b}.

In Refs.~\cite{LRT.07,LRV.07}, we have extended the relativistic RPA by
introducing a coupling to collective vibrations using the techniques developed
and realized long ago for non-relativistic approaches in terms of the Green's
function formalism~\cite{Tse.89a,Tse.89,KTT.97,KST.04}. An induced additional
interaction between single-particle and vibrational excitations provided a
strong fragmentation of the pure RRPA states causing the spreading width of
giant resonances and the redistribution of the pygmy strength to lower
energies. This method does not include pairing correlations and therefore it
is restricted essentially to the few nuclei with doubly closed shells in the
nuclear chart.

In the present work we consider systems with pairing correlations. Again, we
are guided by ideas of the quasiparticle time-blocking approximation (QTBA)
developed and applied for non-relativistic systems in Refs.~\cite{Tse.07} and
\cite{LT.07}, which takes into account quasiparticle-phonon coupling (QPC) and
pairing correlations on an equal footing. However, our approach is based on
CDFT and formulated in terms of relativistic Green's functions of the
Dirac-Hartree-Bogoliubov (DHB) or the Dirac-Hartree-BCS (DHBCS) equations.
Similar, but in details different, approaches developed earlier within a
non-relativistic formalism can be found in Refs.~\cite{Sol.92,BP.99,CB.01}.

The main assumption of the quasiparticle-phonon coupling model~\cite{BM.75} is
that the two types of elementary excitations -- two-quasiparticle and
vibrational modes -- are coupled in such a way that configurations of
$2q\otimes phonon$ type with low-lying phonons strongly compete with simple
$2q$ configurations close in energy or, in other words, that quasiparticles
can emit and absorb phonons with rather high probabilities. Obviously, these
processes should affect both the ground and excited states and therefore, the
corresponding amplitudes should be taken into account both in the
single-nucleon self-energy and in the effective interaction in the nuclear interior.

In order to describe excited states in nuclei, we extend covariant density
functional theory by coupling the quasiparticles to low-lying vibrations in a
consistent way using effective interactions derived from the same Lagrangian
without additional phenomenological parameters. First of all, we use the
well-known quasiparticle formalism, where, in terms of second quantization,
nucleon creation and annihilation operators become components of a
two-component operator mixing a creation and annihilation of a particle into a
single quasiparticle. This leads to the fact, that for systems with pairing
correlations all quantum operators become tensors in the two-dimensional
quasiparticle space. In particular, the relativistic energy functional is
expressed in terms of the relativistic extension of the Valatin density matrix
\cite{Val.61} of double dimension containing the normal as well as the
abnormal densities. As discussed in detail in Refs.
\cite{Kuc.89,KuR.91,SRRi.02,SR.02} pairing correlations can be considered in a
very good approximation as a non-relativistic effect and therefore the full
density functional is a sum of the relativistic energy functional depending on
the normal density and derived from the underlying Lagrangian and a
non-relativistic pairing energy $E_{pair}$, depending on the abnormal density.
The equations of motion are the self-consistent Relativistic Hartree
Bogoliubov (RHB) equations. They are derived from this general functional by
variation with respect to the Valatin density matrix. They are solved
numerically and the self-consistent fields obtained in this way, which do not
depend on the energy, form the static part of the nucleon self-energy. This
static part determines the nuclear ground state in the mean field approximation.

The static effective interaction used in conventional QRPA approximation is
derived as the second derivative of the same energy functional and therefore
it contains no additional parameters. It enables us to go a step further and
to compute amplitudes, or vertices, which describe the emission or absorption
of phonons by quasiparticles within the relativistic framework. These
amplitudes form the essential ingredient for the following considerations.
They determine an additive energy-dependent and non-local term in the
self-energy of the single-quasiparticle equation of motion and, consequently,
an induced effective interaction between the quasiparticles. Both of these
quantities have an influence on the $ph$- as well as on the $pp$-channel.

For the calculation of the response of a nucleus in an external field we use
the Bethe-Salpeter equation. It contains both the static and the induced
effective interactions and it is formulated in the doubled two-quasiparticle
basis of the Dirac-Hartree-Bogoliubov eigenstates. This Bethe-Salpether
equation describes the quasiparticle-phonon coupling and pairing correlations
on the equal footing. It is solved using the quasiparticle time blocking
approximation (QTBA) developed in Ref.~\cite{Tse.07}, which allows the
truncation to $2q\otimes phonon$ configurations and guarantees that the
solution is positive defined. We also use the subtraction procedure introduced
and justified in the Ref.~\cite{Tse.07}. As in the case without pairing it
avoids double counting of the QPC. At zero energy, i.e. at the ground state,
particle vibrational coupling should have no influence, because the
correlations induced by QPC in the ground state have already been taken into
account in the RHB description through the parameters of the energy
functional initially fitted to reproduce experimental data, such as
nuclear binding energies and radii. Therefore, the relativistic mean field
contains effectively all the correlations in the static approximation. The
energy dependence of the self energy influences only excitations at finite
energy in the nucleus.

In the present work we develop the Relativistic Quasiparticle Time Blocking
Approximation (RQTBA) and apply it for the description of electric dipole
excitations in even-even spherical open-shell nuclei, such as the tin ($Z=50$)
isotopes $^{100,106,114,116,120,130}$Sn and the ($N=50$) isotones $^{88}$Sr,
$^{90}$Zr, $^{92}$Mo. The RQTBA method, whose physical content is an extension
of the RQRPA by a coupling to low-lying collective vibrations, provides
spectra enriched with the $2q\otimes phonon$ states. They cause a strong
redistribution of the RQRPA strength. As a result, we obtain an additional
broadening of the giant dipole resonance and a spreading of the soft dipole
mode (pygmy resonance) to lower energies in the nuclei under investigation.

The paper is organized as follows. In Section \ref{form} we formulate basic
relations of our approach in a rather general form. In Section \ref{appl} we
give a more detailed formalism for spherical nuclei in the form adopted for
numerical calculations. Section \ref{details} is devoted to the description of
some numerical details and to the presentation of our results for even-even
semi-magic nuclei. Finally, Section \ref{outlook} contains conclusions and an outlook.

\section{General formalism}

\label{form}

\subsection{Basic relations of the covariant density functional theory for
nuclei with pairing}

In this subsection we recall the general formalism of covariant density
functional theory with pairing, introduce notations and determine conventions
used later on.

In open-shell nuclei, pairing correlations play an essential role and have to
be incorporated consistently in a description of the ground state as well as
of excited states including many-body dynamics. Considering $pp$-correlations
in addition to the usual $ph$-interaction, existing in normal systems, one has
to provide a unified description of both $pp$- and $ph$-channels.

In contrast to Hartree- or Hartree-Fock theory, where $pp$-correlations are
neglected, and where the building blocks of excitations (the quasiparticles in
the sense of Landau) are either nucleons in levels above the Fermi surface
(particles) or missing nucleons in levels below the Fermi surface (holes), we
have now quasiparticles in the sense of Bogoliubov which are described by a
combination of creation and annihilation operators. This fact can be expressed
in a standard way by introducing the following two-component operator, which
is a generalization of the usual particle annihilation operator:
\begin{equation}
\Psi(1)=\left(
\begin{array}
[c]{c}%
a(1)\\
a^{\dagger}(1)
\end{array}
\right)  . \label{psi}%
\end{equation}
Here $a(1)=e^{iHt_{1}}a_{k_{1}}e^{-iHt_{1}}$ is a nucleon annihilation
operator in the Heisenberg picture and the quantum numbers $k_{1}$ represent
an arbitrary basis, $1=\{k_{1},t_{1}\}$. In order to keep the notation simple
we use in the following $1=\{\mbox{\boldmath $r$}_{1},t_{1}\}$ and omit spin
and isospin indices.

Let us introduce the chronologically ordered product of the operator $\Psi(1)$
in Eq. (\ref{psi}) and its Hermitian conjugated operator $\Psi^{\dagger}(2)$,
averaged over the ground state $|\Phi_{0}\rangle$ of the system which will be
concretized below. This tensor of rank 2
\begin{equation}
G(1,2)=-i\langle\Phi_{0}|T\Psi(1)\Psi^{\dagger}(2)|\Phi_{0}\rangle
\ \ \label{fg}%
\end{equation}
is the generalized Green's function which can be expressed through a 2$\times
$2 matrix:
\begin{align}
G(1,2)  &  =-i\theta(t_{1}-t_{2})\langle\Phi_{0}|\left(
\begin{array}
[c]{cc}%
a(1)a^{\dagger}(2) & a(1)a(2)\\
a^{\dagger}(1)a^{\dagger}(2) & a^{\dagger}(1)a(2)
\end{array}
\right)  |\Phi_{0}\rangle\nonumber\\
&  +i\theta(t_{2}-t_{1})\langle\Phi_{0}|\left(
\begin{array}
[c]{cc}%
a^{\dagger}(2)a(1) & a(2)a(1)\\
a^{\dagger}(2)a^{\dagger}(1) & a(2)a^{\dagger}(1)
\end{array}
\right)  |\Phi_{0}\rangle. \label{gmat}%
\end{align}
Similar definitions for the Green's function in non-relativistic superfluid
systems have been used in Refs. \cite{Gor.58,BW.63,Tse.07,LT.07}. Notice that
we define the definition of Green's functions here in the way of
non-relativistic many-body theory, which differs form the conventional
definition $\langle T\Psi\bar{\Psi}\rangle$ adopted in relativistic field
theories by the replacement of $\bar{\Psi}$ by $\Psi^{\dagger}$, i.e. by a
Dirac matrix $\beta=\gamma_{0}$. This notation is more convenient for our
analysis and the matrix $\beta$ needed for Lorentz invariance is included in
the vertices. Therefore the generalized density matrix is obtained as a limit
\begin{equation}
\mathcal{R}(\mbox{\boldmath $r$}_{1},\mbox{\boldmath
$r$}_{2},t_{1})=-i\lim\limits_{t_{2}\rightarrow t_{1}+0}G(1,2) \label{limg}%
\end{equation}
from the second term of Eq. (\ref{gmat}), and, in the notation of Valatin
\cite{Val.61}, it can be expressed as a matrix of doubled dimension containing
as components the normal density ${{\rho}}$ and the abnormal density
${{\varkappa}}$, the so called pairing tensor:
\begin{equation}
\mathcal{R}(\mbox{\boldmath $r$}_{1},\mbox{\boldmath
$r$}_{2},t)=\left(
\begin{array}
[c]{cc}%
\rho(\mbox{\boldmath $r$}_{1},\mbox{\boldmath
$r$}_{2},t) & \varkappa(\mbox{\boldmath $r$}_{1},\mbox{\boldmath
$r$}_{2},t)\\
-\varkappa^{\ast}(\mbox{\boldmath $r$}_{1},\mbox{\boldmath
$r$}_{2},t) & \delta(\mbox{\boldmath $r$}_{1}-\mbox{\boldmath
$r$}_{2})-\rho^{\ast}(\mbox{\boldmath $r$}_{1},\mbox{\boldmath
$r$}_{2},t)
\end{array}
\right)  . \label{rvalatin}%
\end{equation}
These densities play a key role in the description of a superfluid many-body system.

In covariant density functional theory for normal systems the ground state of
the nucleus is a Slater determinant describing nucleons, which move
independently in meson fields $\phi_{m}$ characterized by their quantum
numbers for spin, parity and isospin. In the present investigation we use the
concept of conventional relativistic mean field theory and include the
$\sigma$, $\omega$, $\rho$-meson fields and the electromagnetic field as the
minimal set of fields providing a rather good quantitative description of bulk
and single-particle properties~in the nucleus \cite{Wal.74,SW.86,Rin.96}. This
means that the index $m$ runs over the different types of fields
$m=\{\sigma,\omega,\rho,A\}$. The summation over $m$ implies in particular
scalar products in Minkowski space for the vector fields and in isospace for
the $\rho$-field. In order to obtain a Lorentz invariant theory, these
classical fields $\phi_{m}=\{\sigma,\omega^{\mu},{\vec{\rho}}^{\ \mu},A^{\mu
}\}$ are generated in a self-consistent way by the exchange of virtual
particles, called mesons, and the photon.

Finally the energy depends in the case without pairing correlations on the
normal density matrix ${\rho}$\ and the various fields $\phi_{m}$:
\begin{equation}
E_{RMF}[{{\rho}},\phi]=\text{Tr}[({\mbox{\boldmath $\alpha$}}\mathbf{p}+\beta
m){{\rho}}]+\sum\limits_{m}\Bigl\{\text{Tr}[(\beta\Gamma_{m}\phi_{m}){{\rho}%
}]\pm\int\Bigl[\frac{1}{2}(\mbox{\boldmath $\nabla$}\phi_{m})^{2}+U_{m}%
(\phi)\Bigr]d^{3}r\Bigr \} \label{ERMF}%
\end{equation}
Here we have neglected retardation effects, i.e. time-derivatives of the
fields $\phi_{m}$. The plus sign in Eq.~(\ref{ERMF}) holds for scalar fields
and the minus sign for vector fields. The trace operation implies a sum over
Dirac indices and an integral in coordinate space. ${\mbox{\boldmath
$\alpha$}}$ and $\beta$ are Dirac matrices and the vertices $\Gamma_{m}$ are
given by
\begin{equation}
\Gamma_{\sigma}=g_{\sigma},\ \ \ \ \Gamma_{\omega}^{\mu}=g_{\omega}\gamma
^{\mu},\ \ \ \ {\vec{\Gamma}}_{\rho}^{\ \mu}=g_{\rho}{\vec{\tau}}\gamma^{\mu
},\ \ \ \ \Gamma_{e}^{\mu}=e\frac{(1-\tau_{3})}{2}\gamma^{\mu} \label{gammas}%
\end{equation}
with the corresponding coupling constants $g_{m}$ for the various meson fields
and for the electromagnetic field.

$\bigskip$The quantities $U_{m}(\phi)$ are, in the case of a linear meson
couplings, given by the term
\begin{equation}
U_{m}(\phi)=\frac{1}{2}m_{m}^{2}\phi_{m}^{2} \label{uphi}%
\end{equation}
containing the meson masses $m_{m}$. For non-linear meson couplings, as for
instance for the $\sigma$-meson in the parameter set NL3 we have, as proposed
in Ref.~\cite{BB.77}:
\begin{equation}
U(\sigma)=\frac{1}{2}m_{\sigma}^{2}\sigma^{2}+\frac{g_{2}}{3}\sigma^{3}%
+\frac{g_{3}}{4}\sigma^{4}\;. \label{NL}%
\end{equation}
with two additional coupling constants $g_{2}$ and $g_{3}$.

In superfluid covariant density functional theory the energy is a functional
of the Valatin density $\mathcal{{R}}$ and the fields $\phi_{m}$. Therefore
Relativistic Hartree-Bogoliubov (RHB) theory can be derived from an energy
functional which depends on the normal density ${{\rho}}$ and the abnormal
density ${{\varkappa}}$ as well as on the meson and Coulomb fields $\phi_{m}$.
We use here a density functional of the form
\begin{equation}
E_{RHB}[{{\rho}},{{\varkappa}},{{\varkappa}}^{\ast},\phi]=E_{RMF}[{{\rho}%
},\phi]+E_{pair}[{{\varkappa},{\varkappa}}^{\ast}] \label{ERHB}%
\end{equation}
where the pairing energy is expressed by an effective interaction ${\tilde{V}%
}^{pp}$ in the $pp$-channel:
\begin{equation}
E_{pair}[{{\varkappa},{\varkappa}}^{\ast}]=\frac{1}{4}Tr[{ {\varkappa}}^{\ast
}{\tilde{V}}^{pp}{{\varkappa}}]. \label{Epair}%
\end{equation}
Here and in the following a tilde sign is used to express the static character
of a quantity, i.e. the fact that it does not depend on the energy. Of course,
in Eq. (\ref{ERHB}) we could also use density dependent pairing forces with
$E_{pair}=E_{pair}[{\rho},{{\varkappa}}]$ as it is done for instance in Refs.
\cite{FTT.00,BE.91}. However, in the present investigation we do not consider
this possibility. The effective interaction $\tilde{V}^{pp}$ in the
particle-particle channel is supposed to be independent on the interaction in
the particle-hole channel (see, e.g., Ref.~\cite{PRN.03}) mediated by the
mesons and the electromagnetic fields determined above. Generally, the form of
$\tilde{V}^{pp}$ is restricted only by the conditions of the relativistic
invariance of $E_{pair}$ with respect to the transformations of the abnormal
densities (see Ref.~\cite{CG.99}). In this section, we consider the general
form of $\tilde{V}^{pp}$ as a non-local function in coordinate representation.
In all the applications discussed in Section \ref{appl} we use for $\tilde
{V}^{pp}$ a simple monopole-monopole interaction.

The classical variational principle applied to the energy functional of Eq.
(\ref{ERHB})
\begin{equation}
\delta\int\limits_{t_{1}}^{t_{2}}\Bigl(\langle\Phi_{0}|i\partial_{t}|\Phi
_{0}\rangle-E_{RHB}[{{\rho}},{{\varkappa}},{{\varkappa}}^{\ast},\phi
]\Bigr)dt=0
\end{equation}
leads to the equation of motion for the generalized density matrix
$\mathcal{R}$:
\begin{equation}
i\partial_{t}{{\mathcal{R}}}=[{{\mathcal{H}}}_{RHB}({{\mathcal{R}}%
}),{{\mathcal{R}}}] \label{rmot}%
\end{equation}
with the RHB Hamiltonian
\begin{equation}
{{\mathcal{H}}}_{RHB}=2\frac{\delta E_{RHB}}{\delta\mathcal{R}}=\left(
\begin{array}
[c]{cc}%
h^{\mathcal{D}}-m-\lambda & \Delta\\
-\Delta^{\ast} & -h^{\mathcal{D}\ast}+m+\lambda
\end{array}
\right)  , \label{HRHB}%
\end{equation}
where $\lambda$ is the chemical potential (conted from the continuum limit).
In the static case we find
\begin{equation}
\lbrack{{\mathcal{H}}}_{RHB}({{\mathcal{R}}}),{{\mathcal{R}}}]=0.
\label{rstat}%
\end{equation}
Because of time reversal invariance the currents vanish and we obtain the
single nucleon Dirac Hamiltonian
\begin{equation}
h^{\mathcal{D}}={\mbox{\boldmath $\alpha$}}\mathbf{p}+\beta(m+{\tilde{\Sigma}%
})\ \ \ \label{hd}%
\end{equation}
with the RMF self-energy
\begin{equation}
{\tilde{\Sigma}}(\mbox{\boldmath $r$})=\sum\limits_{m}\Gamma_{m}\phi
_{m}(\mbox{\boldmath $r$}) \label{tsig}%
\end{equation}
The pairing field $\Delta$ reads in this case:
\begin{equation}
\Delta(\mbox{\boldmath $r$},\mbox{\boldmath
$r$}^{\prime})=\frac{1}{2}\int d{\mbox{\boldmath
$r$}}^{\prime\prime}d{\mbox{\boldmath $r$}}^{\prime\prime\prime}{\tilde{V}%
}^{pp}(\mbox{\boldmath $r$},\mbox{\boldmath $r$}^{\prime}%
,\mbox{\boldmath $r$}^{\prime\prime},\mbox{\boldmath
$r$}^{\prime\prime\prime})\varkappa(\mbox{\boldmath
$r$}^{\prime\prime},\mbox{\boldmath $r$}^{\prime\prime\prime}). \label{del}%
\end{equation}

Eq. (\ref{rstat}) leads to the relativistic Hartree-Bogoliubov equations
\cite{KuR.91}
\begin{equation}
\mathcal{H}_{RHB}|\psi_{k}^{\eta}\rangle=\eta E_{k}|\psi_{k}^{\eta}%
\rangle,\ \ \ \ \eta=\pm1 \label{hb}%
\end{equation}
where $|\psi_{k}^{\eta}\rangle$ are the eigenfunctions corresponding to
eigenvalues $\eta E_{k}$. They are the 8-dimensional Bogoliubov-Dirac spinors
of the following form
\begin{equation}
|\psi_{k}^{+}(\mbox{\boldmath $r$})\rangle=\left(
\begin{array}
[c]{c}%
U_{k}(\mbox{\boldmath $r$})\\
V_{k}(\mbox{\boldmath $r$})
\end{array}
\right)  ,\ \ \ \ |\psi_{k}^{-}(\mbox{\boldmath $r$})\rangle=\left(
\begin{array}
[c]{c}%
V_{k}^{\ast}(\mbox{\boldmath $r$})\\
U_{k}^{\ast}(\mbox{\boldmath $r$})
\end{array}
\right)  . \label{dbasis}%
\end{equation}
Note that the index $k$ labels here and in the following quasiparticles in
contrast to the index $k_{1}$ used after Eq. (\ref{psi}) for the particle
basis. In the following we call this quasiparticle basis the
Dirac-Hartree-Bogoliubov (DHB) basis.

The generalized density matrix is obtained as follows:
\begin{equation}
\mathcal{R}(\mbox{\boldmath $r$},\mbox{\boldmath $r$}^{\prime})=\sum
\limits_{k}|\psi_{k}^{-}(\mbox{\boldmath $r$})\rangle\langle\psi_{k}%
^{-}(\mbox{\boldmath $r$}^{\prime})|,
\end{equation}
where the summation is performed only over the states having large upper
components of the Dirac spinors (i.e. large functions $f_{(k)}(r)$ in Eq.
(\ref{sphds}) below). This restriction corresponds to the so-called
\textit{no-sea approximation} (see Ref. \cite{SR.02}).

The behavior of the meson and Coulomb fields is derived from the energy
functional (\ref{ERHB}) by variation with respect to the fields $\phi_{m}$. We
obtain Klein-Gordon equations. In the static case they have the form
\begin{equation}
-\Delta\phi_{m}(\mbox{\boldmath $r$})+U^{\prime}(\phi_{m}%
(\mbox{\boldmath $r$}))=\mp\sum\limits_{k}V_{k}^{\intercal}%
(\mbox{\boldmath $r$})\beta\Gamma_{m}V_{k}^{\ast}(\mbox{\boldmath $r$}),
\label{kg}%
\end{equation}
Eq. (\ref{kg}) determines the potentials entering the single-nucleon Dirac
Hamiltonian (\ref{hd}) and is solved self-consistently together with Eq.
(\ref{hb}). The system of Eqs. (\ref{hb}) and (\ref{kg}) determine the ground
state of an open-shell nucleus in the relativistic Hartree-Bogoliubov approach.

\subsection{Quasiparticle-vibration coupling as a model for an energy
dependence of the single-quasiparticle self-energy}

The single quasiparticle equation of motion (\ref{hb}) determines the behavior
of a nucleon with a static self energy. To include dynamics, i.e. a more
realistic time dependence in the self energy one has to extend the energy
functional by an appropriate term leading to a self-energy (\ref{tsig}) with
time dependence. In the present work we use for this purpose the successful
but relatively simple particle-vibration coupling model introduced in
Refs.~\cite{BM.75,AGD.63}. Following the general logic of this model, we
consider the total single-nucleon self-energy for the Green's function defined
in Eq. (\ref{fg}) as a sum of the RHB self-energy and an energy-dependent
non-local term in the doubled space:
\begin{equation}
{{\Sigma}}(\mbox{\boldmath $r$},\mbox{\boldmath $r$}^{\prime};\varepsilon
)={{\tilde{\Sigma}}}(\mbox{\boldmath $r$},\mbox{\boldmath $r$}^{\prime
})+{{\Sigma}^{(e)}}(\mbox{\boldmath $r$},\mbox{\boldmath $r$}^{\prime
};\varepsilon) \label{Sigma}%
\end{equation}
with
\begin{equation}
{{\tilde{\Sigma}}}(\mbox{\boldmath $r$},\mbox{\boldmath $r$}^{\prime})=\left(
\begin{array}
[c]{cc}%
\beta{\tilde{\Sigma}}(\mbox{\boldmath $r$})\delta
(\mbox{\boldmath $r$}-\mbox{\boldmath $r$}^{\prime}) & \Delta
(\mbox{\boldmath $r$},\mbox{\boldmath $r$}^{\prime})\\
-\Delta^{\ast}(\mbox{\boldmath $r$},\mbox{\boldmath $r$}^{\prime}) &
-\beta{\tilde{\Sigma}}^{\ast}(\mbox{\boldmath $r$})\delta
(\mbox{\boldmath $r$}-\mbox{\boldmath $r$}^{\prime})
\end{array}
\right)  .
\end{equation}
The energy-dependent operator ${{\Sigma}^{(e)}}%
(\mbox{\boldmath $r$},\mbox{\boldmath $r$}^{\prime};\varepsilon)$ will be
determined below (the upper index $e$ in this quantity indicates the energy
dependence). The Dyson equation for the single-quasiparticle Green's
function\ (\ref{fg}) in the doubled space has the following form:
\begin{equation}
(\varepsilon-\mathcal{H}_{RHB}^{{}}-\Sigma^{(e)}(\varepsilon))G(\varepsilon)=1
\label{dyson}%
\end{equation}
To study the influence of the energy-dependent part of the self-energy on the
single-quasiparticle energies, it is convenient to formulate Eq. (\ref{dyson})
in the basis of the eight-component Dirac spinors $|\psi_{k}^{\eta}\rangle$
which diagonalize the static RHB-Hamiltonian $\mathcal{H}_{RHB}$ in Eq.
(\ref{hb}):
\begin{equation}
\sum\limits_{\eta=\pm1}\sum\limits_{k}\Bigl((\varepsilon-\eta_{1}E_{k_{1}%
})\delta_{\eta_{1}\eta}\delta_{k_{1}k}-\Sigma_{k_{1}k}^{(e)\eta_{1}\eta
}(\varepsilon)\Bigr)G_{kk_{2}}^{\eta\eta_{2}}(\varepsilon)=\delta_{\eta
_{1}\eta_{2}}\delta_{k_{1}k_{2}}, \label{fgdb}%
\end{equation}
where
\begin{equation}
\Sigma_{k_{1}k_{2}}^{(e)\eta_{1}\eta_{2}}(\varepsilon)=\int d^{3}rd^{3}%
{r}^{\prime}~\langle{\psi}_{k_{1}}^{\eta_{1}}(\mbox{\boldmath
$r$})|{{\Sigma}}^{(e)}({\mbox{\boldmath $r$}},{\mbox{\boldmath $r$}^{\prime}%
};\varepsilon)|\psi_{k_{2}}^{\eta_{2}}({\mbox{\boldmath $r$}^{\prime}}%
)\rangle, \label{sigmae}%
\end{equation}%
\begin{equation}
G_{k_{1}k_{2}}^{\eta_{1}\eta_{2}}(\varepsilon)=\int d^{3}rd^{3}{r}^{\prime
}~\langle{\psi}_{k_{1}}^{\eta_{1}}(\mbox{\boldmath
$r$})|\,{{G}}({\mbox{\boldmath $r$}},{\mbox{\boldmath
$r$}^{\prime}};\varepsilon)|\psi_{k_{2}}^{\eta_{2}}%
({\mbox{\boldmath $r$}^{\prime}})\rangle. \label{gfme}%
\end{equation}
In this basis the single-quasiparticle Green's function ${\tilde{G}}$ of the
static mean field has the following simple diagonal form:
\begin{equation}
{\tilde{G}}_{k_{1}k_{2}}^{\eta_{1}\eta_{2}}(\varepsilon)=\delta_{k_{1}k_{2}%
}\delta_{\eta_{1}\eta_{2}}{\tilde{G}}_{k_{1}}^{\eta_{1}}(\varepsilon
),\ \ \ \ \ {\tilde{G}}_{k_{1}}^{\eta_{1}}(\varepsilon)=\frac{1}%
{\varepsilon-\eta_{1}E_{k_{1}}+i\eta_{1}\delta},\ \ \ \ \ \ \delta
\rightarrow+0. \label{mfgf}%
\end{equation}

As in Refs.~\cite{LR.06,LRT.07}, we use the particle-phonon coupling model for
the energy-dependent part of the self-energy $\Sigma^{(e)}$. In the basis of
the spinors $|\psi_{k}^{\eta}\rangle$ of Eq. (\ref{dbasis}), called in the
following \textit{Dirac basis}, its matrix elements are given by:
\begin{equation}
\Sigma_{k_{1}k_{2}}^{(e)\eta_{1}\eta_{2}}(\varepsilon)=\sum\limits_{\eta=\pm
1}\sum\limits_{\eta_{\mu}=\pm1}\sum\limits_{k,\mu}\frac{\delta_{\eta_{\mu
},\eta_{3}}{\gamma}_{\mu;k_{1}k}^{\eta_{\mu};\eta_{1}\eta}\ {\gamma}%
_{\mu;k_{2}k}^{\eta_{\mu};\eta_{2}\eta\ast}}{\varepsilon-\eta E_{k}-\eta_{\mu
}(\Omega_{\mu}-i\delta)},\ \ \ \ \delta\rightarrow+0. \label{sgephon}%
\end{equation}
The index $k_{3}$ formally runs over all single-quasiparticle states in the
DHB basis including antiparticle states with negative energies. In the doubled
quasiparticle space we can no longer distinguish occupied and unoccupied
states considering that all the orbits are partially occupied. But in
practical calculations, it is assumed that there are no pairing correlations
in the Dirac sea \cite{SR.02} and the orbits with negative energies are
treated in the no-sea approximation. As it has been shown in calculations for
nuclei with closed shells in Ref.~\cite{LR.06}, the numerical contribution of
the diagrams with intermediate states $k$ with negative energy is very small
due to the large energy denominators in the corresponding terms of the
self-energy (\ref{sgephon}). The index $\mu$ in Eq.~(\ref{sgephon}) labels the
set of phonons taken into account. $\Omega_{\mu}$ are their frequencies and
$\eta_{\mu}=\pm1$ labels forward and backward going diagrams in Eq.
(\ref{sgephon}). The vertices ${\gamma}_{\mu;k_{1}k_{2}}^{\eta_{\mu};\eta
_{1}\eta_{2}}$ determine the coupling of the quasiparticles, to the collective
state $\mu$:
\begin{equation}
{\gamma}_{\mu;k_{1}k_{2}}^{\eta_{\mu};\eta_{1}\eta_{2}}=\delta_{\eta_{\mu}%
,+1}{\gamma}_{\mu;k_{1}k_{2}}^{\eta_{1}\eta_{2}}+\delta_{\eta_{\mu},-1}%
{\gamma}_{\mu;k_{2}k_{1}}^{\eta_{2}\eta_{1}\ast},
\end{equation}
In the conventional version of the particle-vibrational coupling model the
phonon vertices $\gamma_{\mu}$ are derived from the corresponding transition
densities $\mathcal{R}_{\mu}$ and the static effective interaction:
\begin{equation}
\gamma_{\mu;k_{1}k_{2}}^{\eta_{1}\eta_{2}}=\sum\limits_{k_{3}k_{4}}%
\sum\limits_{\eta_{3}\eta_{4}}\tilde{V}_{k_{1}k_{4},k_{2}k_{3}}^{\eta_{1}%
\eta_{4},\eta_{2}\eta_{3}}\mathcal{R}_{\mu;k_{3}k_{4}}^{\eta_{3}\eta_{4}},
\label{phonon}%
\end{equation}
where $\tilde{V}_{k_{1}k_{4},k_{2}k_{3}}^{\eta_{1}\eta_{4},\eta_{2}\eta_{3}}$
denotes a relativistic matrix element of the static residual interaction in
the doubled space. It is obtained as a functional derivative of the
relativistic mean-field self-energy ${{\tilde{\Sigma}}}$ with respect to the
relativistic generalized density matrix ${\mathcal{R}}$:
\begin{equation}
{\tilde{V}}_{k_{1}k_{4},k_{2}k_{3}}^{\eta_{1}\eta_{4},\eta_{2}\eta_{3}}%
=\frac{\ \delta{\tilde{\Sigma}}_{k_{4}k_{3}}^{\eta_{4}\eta_{3}}\ }%
{\ \delta\mathcal{R}_{k_{2}k_{1}}^{\eta_{2}\eta_{1}}\ }.
\label{static-interaction}%
\end{equation}
The transition densities $\mathcal{R}_{\mu}$ are defined by the time
dependence of the generalized density (\ref{rvalatin})%
\begin{equation}
\mathcal{R}(t)=\mathcal{R}_{0}+%
{\displaystyle\sum\limits_{\mu}}
(\mathcal{R}_{\mu}e^{i\Omega_{\mu}t}+\text{h.c.)} \label{rhomu}%
\end{equation}
describing the oscillating system. We use the linearized version of the model
which assumes that the transition densities $\mathcal{R}_{\mu}$ are not
influenced by the particle-phonon coupling and that they can be computed
within relativistic QRPA. In the linearized version of the QPC model we solve
the usual QRPA equations for transition densities
\begin{equation}
\mathcal{R}_{\mu;k_{1}k_{2}}^{\eta}={\tilde{R}}_{k_{1}k_{2}}^{(0)\eta}%
(\Omega_{\mu})\sum\limits_{k_{3}k_{4}}\sum\limits_{\eta^{\prime}}\tilde
{V}_{k_{1}k_{4},k_{2}k_{3}}^{\eta\eta^{\prime}}\mathcal{R}_{\mu;k_{3}k_{4}%
}^{\eta^{\prime}} \label{qrpa}%
\end{equation}
where
\begin{equation}
\mathcal{R}_{\mu;k_{1}k_{2}}^{\eta}=\mathcal{R}_{\mu;k_{1}k_{2}}^{\eta,-\eta
},\text{ \ \ \ \ \ \ \ \ }{\tilde{R}}_{k_{1}k_{2}}^{(0)\eta}(\omega
)={\tilde{R}}_{k_{1}k_{2}}^{(0)\eta,-\eta}(\omega),\text{ \ \ \ \ \ \ }%
\tilde{V}_{k_{1}k_{4},k_{2}k_{3}}^{\eta\eta^{\prime}}=\tilde{V}_{k_{1}%
k_{4},k_{2}k_{3}}^{\eta,-\eta^{\prime},-\eta,\eta^{\prime}},
\end{equation}
which means that we cut out certain components of the tensors in the
quasiparticle space. The quantity ${\tilde{R}}$ is, as usual, the
two-quasiparticle propagator, or the mean-field response function, which is a
convolution of two single-quasiparticle mean-field Green's functions
(\ref{mfgf}):
\begin{equation}
{\tilde{R}}_{k_{1}k_{2}}^{(0)\eta}(\omega)=\frac{1}{\eta\omega-E_{k_{1}%
}-E_{k_{2}}}.
\end{equation}
In Eq. (\ref{qrpa}) we use the static quasiparticle-interaction ${\tilde{V}}$
$\ $\ of Eq. (\ref{static-interaction}). Of course, in general, we should
calculate these transition densities taking into account the also the
additional energy-dependent residual interaction ${V}^{(e)}$ [see
Eq.~(\ref{ueampl}) below]\textsl{ }
in a self-consistent iteration procedure. However, this is not done in the
investigations presented here.

\subsection{Response function in the quasiparticle time-blocking
approximation}

Now we have to formulate the Bethe-Salpeter equation (BSE) for the response of
a superfluid nucleus in a weak external field. The method to derive the BSE
for superfluid non-relativistic systems from a generating functional is known
and can be found, e.g.,
in Ref. \cite{Tse.07} where the generalized Green's function formalism was
used. Applying the same technique in the relativistic case, one obtains a
similar ansatz for the BSE. It is formulated now in the basis of the DHB
spinors in Eq. (\ref{dbasis}). In full analogy to the case without pairing
described in Ref. \cite{LRT.07} it is convenient to begin in the time
representation. Let us therefore include the time variable and the variable
$\eta$ defined in Eq. (\ref{hb}), which distinguishes components in the
doubled quasiparticle space, into the single-quasiparticle indices using
$1=\{k_{1},\eta_{1},t_{1}\}$. In this notation the BSE for the response
function $R$ reads:
\begin{equation}
R(14,23)=G(1,3)G(4,2)-i\sum\limits_{5678}G(1,5)G(6,2)V(58,67)R(74,83),
\label{bse0}%
\end{equation}
where the summation over the number indices $1$, $2,\dots$ implies integration
over the respective time variables. The function $G$ is the exact
single-quasiparticle Green's function, and $V$ is the amplitude of the
effective interaction irreducible in the $ph$-channel. This amplitude is
determined as a variational derivative of the full self-energy $\Sigma$ with
respect to the exact single-quasiparticle Green's function:
\begin{equation}
V(14,23)=i\frac{\delta\Sigma(4,3)}{\delta G(2,1)}. \label{uampl}%
\end{equation}
Similar as in Ref. \cite{LRT.07}, we introduce the free response
$R^{0}(14,23)=G(1,3)G(4,2)$ and formulate the Bethe-Salpeter equation
(\ref{bse0}) in a shorthand notation, omitting the number indices:
\begin{equation}
R=R^{0}-iR^{0}VR. \label{bse0s}%
\end{equation}
For the sake of simplicity, we will use this shorthand notation in the
following discussions. Since the self-energy in Eq.~(\ref{Sigma}) has two
parts ${\Sigma}={\tilde{\Sigma}}+{\Sigma}^{(e)}$, the effective interaction
${V}$ in Eq.~(\ref{bse0}) is a sum of the static RMF interaction ${\tilde{V}}$
and the energy-dependent term ${V}^{(e)}$:
\begin{equation}
V={\tilde{V}}+{V}^{(e)},
\end{equation}
where (with $t_{12}=t_{1}-t_{2}$)
\begin{equation}
\tilde{V}(14,23)=\tilde{V}_{k_{1}k_{4},k_{2}k_{3}}^{\eta_{1}\eta_{4},\eta
_{2}\eta_{3}}\delta(t_{31})\delta(t_{21})\delta(t_{34})\,, \label{V-static}%
\end{equation}%
\begin{equation}
V^{(e)}(14,23)=i\frac{\delta\Sigma^{(e)}(4,3)}{\delta G(2,1)}\,,
\label{e-interaction}%
\end{equation}
and\textsl{ $\tilde{V}_{k_{1}k_{4},k_{2}k_{3}}^{\eta_{1}\eta_{4},\eta_{2}%
\eta_{3}}$ }is determined by Eq\textsl{.~}(\ref{static-interaction})\textsl{.
}
In the DHB basis of Eq. (\ref{dbasis}) the Fourier transform of the amplitude
$V^{(e)}$ has the form:
\begin{equation}
V_{k_{1}k_{4},k_{2}k_{3}}^{(e)\eta_{1}\eta_{4},\eta_{2}\eta_{3}}%
(\omega,\varepsilon,\varepsilon^{\prime})=\sum\limits_{\mu,\eta_{\mu}}%
\frac{\eta_{\mu}\gamma_{\mu;k_{3}k_{1}}^{\eta_{\mu};\eta_{3}\eta_{1}}%
\gamma_{\mu;k_{4}k_{2}}^{\eta_{\mu};\eta_{4}\eta_{2}\ast}}{\varepsilon
-\varepsilon^{\prime}+\eta_{\mu}(\Omega_{\mu}-i\delta)}\ ,\ \ \ \ \ \ \delta
\rightarrow+0\ . \label{ueampl}%
\end{equation}
In order to make the Bethe-Salpeter equation (\ref{bse0s}) more convenient for
the further analysis we eliminate the exact Green's function $G$ and rewrite
it in terms of the mean field Green's function $\tilde{G}$ which is diagonal
in the DHB basis. In time representation it has the following ansatz:
\begin{equation}
{\tilde{G}}(1,2)=-i\eta_{1}\delta_{k_{1}k_{2}}\theta(\eta_{1}\tau
)e^{-i\eta_{1}E_{k_{1}}\tau},\ \ \ \ \ \tau=t_{1}-t_{2},
\end{equation}
and its Fourier transform is given by Eq. (\ref{mfgf}).

Using the connection between the mean field GF $\tilde{G}$ and the exact GF
$G$ in the Nambu form
\begin{equation}
{\tilde{G}}^{-1}(1,2)=G^{-1}(1,2)+\Sigma^{e}(1,2),
\end{equation}
one can eliminate the unknown exact GF $G$ from the Eq.~(\ref{bse0s}) and
rewrite it as follows:
\begin{equation}
R={\tilde{R}}^{0}-i\tilde{R}^{0}WR \label{bse}%
\end{equation}
with the mean-field response ${\tilde{R}}^{0}(14,23)={\tilde{G}}%
(1,3){\tilde{G}}(4,2)$, and $W$ is a new interaction of the form
\begin{equation}
W=\tilde{V}+W^{(e)}, \label{wampl}%
\end{equation}
where
\begin{equation}
W^{(e)}(14,23)=V^{(e)}(14,23)+i\Sigma^{(e)}(1,3){\tilde{G}}^{-1}%
(4,2)+i{\tilde{G}}^{-1}(1,3)\Sigma^{(e)}(4,2)-i\Sigma^{(e)}(1,3)\Sigma
^{(e)}(4,2). \label{wampl1}%
\end{equation}
Thus, we have obtained the BSE in terms of the mean-field propagator,
containing the well-known mean-field Green's functions ${\tilde{G}}$, and a
rather complicated effective interaction $W$ in Eq.~(\ref{wampl}), which,
however, is also expressed through the mean-field Green's functions.

The structure of the energy-dependent effective interaction $W^{(e)}$ has a
clear interpretation in terms of Feynman's diagrams which are usually employed
to clarify the physical content of the amplitude $W^{(e)}$
\cite{Tse.89,Tse.07}. In addition to the static interaction $\tilde{V}$, the
effective interaction $W$ contains diagrams with energy-dependent
self-energies and an energy-dependent induced interaction, where a phonon is
exchanged between the two quasiparticles. In the present work, as well as in
Ref. \cite{LRT.07}, we omit the term $i\Sigma^{(e)}(3,1)\Sigma^{(e)}(2,4)$ in
Eq.~(\ref{wampl1}) because it plays a compensational role with respect to the
backward-going components of the previous terms in the $W^{(e)}$. However,
within the version of\textsl{ } the time blocking approximation, which we
apply to the BSE (see below), the backward-going propagators are not taken
into account. Components containing the backward-going propagators within
$2q\otimes phonon$ configurations require a special consideration which is
formulated in Ref. \cite{Tse.07} for a superfluid non-relativistic system. In
the present work these correlations are fully neglected and, therefore, the
term $i\Sigma^{(e)}(3,1)\Sigma^{(e)}(2,4)$ has also to be omitted. However, we
have to emphasize, that we only neglect ground state correlations (GSC)
(backward-going diagrams) caused by the quasiparticle-phonon coupling. All the
QRPA ground state correlations are taken into account, because it is well
known that they play a central role for the conservation of currents and sum
rules. We consider that this is a reasonable approximation which is applied
and discussed also in some non-relativistic models (see e.g. Refs.
\cite{SSV.77,BBBD.79,BB.81,CBG.92,CB.01,SBC.04,Tse.89,KTT.97,KST.04,Tse.07,LT.07}
and references therein).

Eq. (\ref{bse}) whose integral part contains singularities in the amplitude
$W$ can not be solved explicitly because, considering the Fourier transform of
the Eq. (\ref{bse}), one finds that both the solution of this equation $R$ and
its kernel $W$ are singular with respect to energy variables. Also, this
equation contains integrations over all time points of the intermediate
states. This implies that many configurations which are actually more complex
than $2q\otimes phonon$ are contained in the exact response function.
Therefore, we apply the special time-projection technique, introduced in the
Ref.~\cite{Tse.89} and generalized in Ref. \cite{Tse.07} for superfluid
systems, to block the $2q$-propagation through these complicated intermediate states.

Conventionally, we divide the problem to find the exact response function of
the BSE (\ref{bse}) into two parts. First, we calculate the correlated
propagator $R^{(e)}$ which describes the $2q$-propagation under the influence of
the interaction $W^{e}$
\begin{equation}
R^{(e)}={\tilde{R}}^{0}-i{\tilde{R}}^{0}W^{(e)}R^{(e)}. \label{RE}%
\end{equation}
It contains all the effects of particle-phonon coupling and all the
singularities of the integral part of the initial BSE. Second, we have to
solve the remaining equation for the full response function $R$
\begin{equation}
R = R^{(e)} - iR^{(e)}\tilde{V}R. \label{respre}%
\end{equation}
Eq.~(\ref{respre}) contains only the static effective interaction $\tilde{V}$
and can be easily solved when $R^{(e)}$ is known.

The correlated propagator $R^{(e)}$ can be represented as an infinite series
of graphs which contain mean-field $2q$-propagators alternated with single
interaction acts. This can be expressed by the system of the following
equations employing the auxiliary amplitude ${\Gamma^{(e)}}$:
\begin{align}
R^{(e)}  &  ={\tilde{R}}^{0}-i{\tilde{R}}^{0}{\Gamma}^{(e)}{\tilde{R}}%
^{0},\label{re}\\
{\Gamma}^{(e)}  &  =W^{(e)}-iW^{(e)}{\tilde{R}}^{0}{\Gamma}^{(e)}.
\label{gamma}%
\end{align}
Then, the integral part of Eq.~(\ref{gamma}) has to be modified to order in
time the interaction acts described by the amplitude $W^{(e)}$. It means that
we should cut out only terms where the 'left' time arguments of the amplitude
$\Gamma^{(e)}$ are greater than the 'right' time arguments of the amplitude
$W^{(e)}$ \cite{Tse.89,Tse.07}. This can be expressed by the time-projection
operator of the form:
\begin{equation}
\Theta(14,23)=\delta_{\eta_{1},-\eta_{2}}\delta_{k_{1}k_{3}}\delta_{k_{2}%
k_{4}}\theta(\eta_{1}t_{14})\theta(\eta_{1}t_{23}), \label{theta}%
\end{equation}
which is introduced into the integral part of the Eq.~(\ref{gamma}):
\begin{equation}
{\Gamma}^{(e)}(14,23)=W^{(e)}(14,23)+\frac{1}{i}\sum\limits_{5678}%
W^{(e)}(16,25){\tilde{R}}^{0}(58,67)\Theta(58,67){\Gamma}^{(e)}(74,83).
\label{gamma1}%
\end{equation}
Since we are interested in spectral characteristics of the nuclear response, a
Fourier transformation of the response function
is performed as follows:
\begin{equation}
R_{k_{1}k_{4},k_{2}k_{3}}^{\eta_{1}\eta_{4},\eta_{2}\eta_{3}}(\omega
)=-i\int\limits_{-\infty}^{\infty}dt_{1}dt_{2}dt_{3}dt_{4}\delta(t_{1}%
-t_{2})\delta(t_{3}-t_{4})\delta(t_{4})e^{i\omega t_{13}}R(14,23),
\label{tbaresp}%
\end{equation}
so that the response function depends only on one energy variable $\omega$.

The time projection by the operator (\ref{theta}) leads, after some algebra
and the transformation (\ref{tbaresp}), to an algebraic equation for the
response function. For the $ph$-type components of the response function it
has the form:
\begin{equation}
R_{k_{1}k_{4},k_{2}k_{3}}^{\eta\eta^{\prime}}(\omega)=\tilde{R}_{k_{1}k_{2}%
}^{(0)\eta}(\omega)\delta_{k_{1}k_{3}}\delta_{k_{2}k_{4}}\delta_{\eta
\eta^{\prime}}+\tilde{R}_{k_{1}k_{2}}^{(0)\eta}(\omega)\sum\limits_{k_{5}%
k_{6}k_{7}k_{8}}\sum\limits_{\eta^{\prime\prime}}{\bar{W}}_{k_{5}k_{8}%
,k_{6}k_{7}}^{\eta\eta^{\prime\prime}}(\omega)R_{k_{7}k_{4},k_{8}k_{3}}%
^{\eta^{\prime\prime}\eta^{\prime}}(\omega), \label{respdir}%
\end{equation}
where
\begin{equation}
{\bar{W}}_{k_{1}k_{4},k_{2}k_{3}}^{\eta\eta^{\prime}}(\omega)=\tilde{V}%
_{k_{1}k_{4},k_{2}k_{3}}^{\eta\eta^{\prime}}+\Bigl(\Phi_{k_{1}k_{4},k_{2}%
k_{3}}^{\eta}(\omega)-\Phi_{k_{1}k_{4},k_{2}k_{3}}^{\eta}(0)\Bigr)\delta
_{\eta\eta^{\prime}}. \label{W-omega}%
\end{equation}
$\Phi$ is the particle-phonon coupling amplitude in the QTBA with the
following forward ($\eta= 1$) and backward ($\eta= -1$) components:
\begin{align}
\Phi_{k_{1}k_{4},k_{2}k_{3}}^{\eta}(\omega)  & = \sum\limits_{\mu}
\Bigl[\delta_{k_{1}k_{3}}\sum\limits_{k_{6}} \frac{\gamma_{\mu;k_{6}k_{2}%
}^{-\eta} \gamma_{\mu;k_{6}k_{4}}^{-\eta\ast}}{\eta\omega-E_{k_{1}}-E_{k_{6}%
}-\Omega_{\mu}} + \delta_{k_{2}k_{4}}\sum\limits_{k_{5}}\frac{\gamma
_{\mu;k_{1}k_{5}}^{\eta} \gamma_{\mu;k_{3}k_{5}}^{\eta\ast}}{\eta\omega-
E_{k_{5}} - E_{k_{2}} - \Omega_{\mu}}\nonumber\\
& \qquad\qquad-\Bigl(\frac{\gamma_{\mu;k_{1}k_{3}}^{\eta} \gamma_{\mu
;k_{2}k_{4}}^{-\eta\ast}}{\eta\omega- E_{k_{3}}- E_{k_{2}} - \Omega_{\mu}} +
\frac{\gamma_{\mu;k_{3}k_{1}}^{\eta\ast}\gamma_{\mu;k_{4}k_{2}}^{-\eta}}
{\eta\omega- E_{k_{1}} - E_{k_{4}} - \Omega_{\mu}}\Bigr)\Bigr],
\label{phiphc0}%
\end{align}
where we denote:
\begin{equation}
\gamma_{\mu;k_{1}k_{2}}^{\eta} = \gamma_{\mu;k_{1}k_{2}}^{\eta\eta}%
\qquad\mbox{and} \qquad\Phi_{k_{1}k_{4},k_{2}k_{3}}^{\eta}(\omega) =
\Phi_{k_{1}k_{4},k_{2}k_{3}}^{\eta,-\eta,-\eta,\eta}(\omega) . \label{phieta}%
\end{equation}
Indices $k_{i}$ in this expression formally run over the whole DHB space, but
in applications we usually consider that the amplitude $\Phi_{k_{1}k_{4}%
,k_{2}k_{3}}^{\eta}(\omega)$ describes phonon coupling only within some energy
window around the Fermi surface. That is why it implies that this amplitude
contains no antiparticle-quasiparticle ($\alpha q$) configurations. Notice,
that in our approach we cut out only the components without ground state
correlations induced by phonon coupling (\ref{phieta}) which include the main
contribution of the phonon coupling and neglect some more delicate terms.

However, ground state correlations of the QRPA type are taken into account due
to the presence of the $\tilde{V}_{k_{1}k_{4},k_{2}k_{3}}^{\eta\eta^{\prime}}$
terms of the static interaction in the Eq.~(\ref{respdir}). By definition, the
propagator $R(\omega)$ in Eq.~(\ref{respdir}) contains only configurations
which are not more complicated than $2q\otimes phonon$.

In Eq. ~(\ref{respdir}) we have included the subtraction procedure because of
the same reasons as in the Ref. \cite{LRT.07}. Since the RMF ground state is
adjusted to experimental data, it contains effectively many correlations in
the static approximation and, in particular, also admixtures of phonons.
Therefore, when we include them explicitly in the dynamics, this static part
should be subtracted from the effective interaction to avoid double counting
of the QPC correlations. Since the parameters of the density functional and,
as a consequence, the effective interaction $\tilde{V}$ are adjusted to
experimental ground state properties at the energy $\omega=0$, this part of
the interaction $\Phi(\omega)$, which is already contained in $\tilde{V}$, is
given by $\Phi(0)$. This subtraction method has been introduced in the
Ref.~\cite{Tse.07} for self-consistent schemes.

Eventually, to describe the observed spectrum of the excited nucleus in a weak
external field $P$ as, for instance, an electromagnetic field, one needs to
calculate the strength function:
\begin{equation}
S(E)=-\frac{1}{\pi}\lim\limits_{\Delta\rightarrow+0}Im\ \Pi(E+i\Delta),
\label{strf}%
\end{equation}
expressed through the polarizability $\Pi(\omega)$ defined as
\begin{equation}
\Pi(\omega)=\frac{1}{2}P^{\dag}R(\omega)P:=\frac{1}{2}\sum\limits_{k_{1}%
k_{2}k_{3}k_{4}}\sum\limits_{\eta\eta^{\prime}}P_{k_{1}k_{2}}^{\eta\ast
}R_{k_{1}k_{4},k_{2}k_{3}}^{\eta\eta^{\prime}}(\omega)P_{k_{3}k_{4}}%
^{\eta^{\prime}}. \label{polarization}%
\end{equation}
The imaginary part $\Delta$ of the energy variable is introduced for
convenience in order to obtain a more smoothed envelope of the spectrum. This
parameter has the meaning of an additional artificial width for each
excitation. This width emulates effectively contributions from configurations
which are not taken into account explicitly in our approach.

In relativistic RPA and QRPA calculations the Dirac sea plays an important
role. A consistent derivation of relativistic RPA (QRPA) as the small
amplitude limit of time-dependent RMF (RHB) theory in Ref. \cite{RMG.01} shows
that one has to include besides the usual $ph$-configurations also
antiparticle-hole ($\alpha h$) configurations. Otherwise current conservation
is violated \cite{DF.90} and the position of giant resonances cannot be
described properly in relativistic RPA \cite{MGT.97}. However, this increases
the number of configurations dramatically as compared to non-relativistic QRPA
calculations and requires in particular in deformed relativistic QRPA
calculations \cite{PR.07} a tremendous large numerical effort. Recently a
simple method has been proposed to avoid this problem. As discussed in Ref.
\cite{SNS}, the \textit{static no-sea} (SNS) approximation takes the
contributions of the empty Dirac sea into account in a very good approximation
by a renormalization of the total effective interaction $\bar{W}(\omega)$ in
the Bethe-Salpeter equation.

\section{Application of the approach: basic approximations}

\label{appl}

The formulated relativistic QTBA is applied to calculations of the dipole
strength in spherical nuclei with pairing. In this application we mainly
follow the calculation scheme employed in Ref. \cite{LRT.07,LRV.07}, however,
with some considerable modifications accounting pairing effects: all the
equations are solved in the doubled space. The computation is performed by the
following main steps:

i) To calculate ground state properties the Dirac equation together with the
BCS equation for single nucleons are solved simultaneously with the
Klein-Gordon equation for meson fields in a self-consistent way to obtain the
single-quasiparticle basis (Dirac-Hartree-BCS basis).

ii) The RQRPA equations (\ref{qrpa}) with the static interaction $\tilde{V}$
of Eq. (\ref{static-interaction}) are solved in the Dirac-Hartree-BCS basis to
determine the low-lying collective vibrations (phonons), their energies and
amplitudes. In the present work we have included the phonon modes with
energies below the neutron separation energies for the Z=50 chain and with
energies below 10 MeV for the N=50 chain. The two sets of quasiparticles and
phonons form the multitude of $2q\otimes phonon$ configurations which enter the
quasiparticle-phonon coupling amplitude $\Phi(\omega)$ in Eq. (\ref{phiphc0}).

iii) The equation for the response function (\ref{bse}) is solved using this
additional amplitude in the effective interaction ${W}(\omega)$. Making a
double convolution of the response function with the external field operator
$P$, one obtains the polarizability (\ref{polarization}) and the strength
function (\ref{strf}) determining the spectrum of the nucleus. It is found
that the amplitude ${\bar{W}}(\omega)$, containing a large number of poles of
$2q\otimes phonon$ nature, provides a considerable enrichment of the calculated
spectrum as compared to the pure RQRPA.

\subsection{Description of the ground state}

In the present work we confine ourselves by the case of spherically symmetric
nuclei where it is convenient to separate the dependence on the magnetic
quantum number $m_{k}$: $k=\{(k),m_{k}\}$, where $(k)$ is the set of remaining
quantum numbers which are time reversal invariant: ${-k}=\{(k),-m_{k}\}$. In
this case $(k)=\{n_{k},j_{k},\pi_{k},{\tau}_{k}\}$ with the radial quantum
number $n_{k}$, angular momentum quantum number $j_{k}$, parity $\pi_{k}$ and
isospin ${\tau}_{k}$, so the Dirac spinors read:
%
\begin{equation}
\varphi_{k}(\mathbf{r},t)=\left(
\begin{array}
[c]{c}%
f_{(k)}(r)\,\mathcal{Y}_{l_{k}j_{k}m_{k}}(\vartheta,\varphi)\\
ig_{(k)}(r)\,\mathcal{Y}_{{\tilde{l}}_{k}j_{k}m_{k}}(\vartheta,\varphi)
\end{array}
\right)  \chi_{{\tau}_{k}}(t), \label{sphds}%
\end{equation}
$\mathcal{Y}_{ljm}(\vartheta,\varphi)$ is a two-component spinor
\begin{equation}
\mathcal{Y}_{ljm}(\vartheta,\varphi,s)=\sum\limits_{m_{s}m_{l}}({\frac
{{\scriptstyle1}}{{\scriptstyle2}}}m_{s}lm_{l}|jm)Y_{lm_{l}}(\vartheta
,\varphi)\chi_{m_{s}}(s)\,,
\end{equation}
$t$ is the coordinate for the isospin and $\chi_{{\tau}_{k}}(t)$ is a spinor
in the isospin space. The orbital angular momenta $l_{k}$ and $\tilde{l}_{k}$
of the large and small components are determined by the parity of the state
$k$:
\begin{equation}
\left\{
\begin{array}
[c]{ccc}%
l_{k}=j_{k}+\frac{1}{2}, & {\tilde{l}}_{k}=j_{k}-\frac{1}{2} &
\mbox{for}\ \ \pi_{k}=(-1)^{j_{k}+\frac{1}{2}}\\
l_{k}=j_{k}-\frac{1}{2}, & {\tilde{l}}_{k}=j_{k}+\frac{1}{2} &
\mbox{for}\ \ \pi_{k}=(-1)^{j_{k}-\frac{1}{2}},
\end{array}
\right.
\end{equation}
$f_{(k)}(r)$ and $g_{(k)}(r)$ are radial wave functions. The phase convention
for the wave function $\varphi_{-k}$ is chosen so that the following relation
is fulfilled:
%
\begin{equation}
\gamma^{3}\gamma^{1}\varphi_{k}^{\ast}=(-)^{l_{k}%
+j_{k}-m_{k}}\varphi_{-k}^{{}}\,. \label{pc}%
\end{equation}

In the literature \cite{PRN.03} the RQRPA are solved for finite range Gogny
forces in the pairing channel in the canonical basis. This has the advantage,
that the quasiparticle matrix elements of the QRPA-equations can be calculated
rather easily by multiplying the matrix elements in particle space by
BCS-occupation factors, but it has the disadvantage, that the matrix $H^{11}$
in quasiparticle space is no longer diagonal in the canonical basis. The
quasiparticle energies $E_{k_{1}}+E_{k_{2}}$ have to be replaced by
complicated matrices.

We therefore use in the following applications the RMF+BCS approximation,
where the canonical basis coincides with the BCS-basis. In this approximation
the ground state wave function $|\Phi_{0}\rangle$ is considered to be a vacuum
state with respect to quasiparticles with the creation and annihilation
operators $\alpha_{k}^{\dagger}$, $\alpha_{k}$ determined by the special
Bogoliubov transformation:
\begin{equation}
\left(
\begin{array}
[c]{c}%
\alpha_{k}\\
\alpha_{\bar{k}}^{\dagger}%
\end{array}
\right)  =\left(
\begin{array}
[c]{cc}%
u_{k} & -v_{k}\\
v_{k} & u_{k}%
\end{array}
\right)  \left(
\begin{array}
[c]{c}%
a_{k}\\
a_{\bar{k}}^{\dagger}%
\end{array}
\right)  ,\ \ \ \ \ \alpha_{k}|\Phi_{0}\rangle=0\ \ \ \forall k, \label{bog}%
\end{equation}
where $u_{k}^{2}+v_{k}^{2}=1.$ Operation ${\bar{k}}$ transforms the state $k$
to the time reversal state. In a spherical system we define
\begin{equation}
a_{{\bar{k}}}=(-1)^{l_{k}+j_{k}-m_{k}}a_{-k},
\end{equation}
where the choice of the phase factors is determined by Eq. (\ref{pc}).

In the RMF+BCS approximation we determine, in each step of the iteration,
first the eigen functions $\varphi_{k}$ of the single-particle Dirac
Hamiltonian $h^{\mathcal{D}}$ of Eq. (\ref{hd})
\begin{equation}
\int dx^{\prime}h^{\mathcal{D}}(x,x^{\prime})\varphi_{k}(x^{\prime
})=(m+\varepsilon_{k})\varphi_{k}(x) \label{dhd}%
\end{equation}
where the coordinate $x=\{{\mbox{\boldmath $r$}},\alpha,t\}$ combines the
spatial coordinates ${\mbox{\boldmath $r$}}$\textbf{ }with the Dirac index
$\alpha=1\dots4$ and the isospin $t$. Next the Dirac spinors $\varphi_{k}$ are
used to construct the single-particle density matrix
\begin{equation}
\rho(x,x^{\prime})=%
{\displaystyle\sum\limits_{k}}
\varphi_{k}(x)v_{k}^{2}\varphi_{k}^{\dagger}(x^{\prime}). \label{rho}%
\end{equation}
In the basis of the functions $\varphi_{k}$ (BCS basis) $\rho$ as well as
$h^{\mathcal{D}}$ are diagonal with the eigenvalues $v_{k}^{2}$ and
$m+\varepsilon_{k}$. The pairing field $\Delta$ is in this basis close to
canonical form: $\Delta_{k{\bar{k}}^{\prime}}=\delta_{kk^{\prime}}\Delta_{k}$.
All other matrix elements vanish in the case of a monopole force with constant
matrix elements and without cut-off, in other cases they are neglected in the
BCS approximation. Thus, in this basis, the Hartree-Bogoliubov
matrix\ (\ref{hb}) is reduced to a set of 2x2 matrices, which can be
diagonalized analytically. Thus one finds as eigenvalues the quasiparticle
energies%
\begin{equation}
\ E_{k}=\sqrt{(\varepsilon_{k}-\lambda_{\tau_{k}})^{2}+\Delta_{k}^{2}}%
\end{equation}
and as eigen functions the occupation amplitudes $u_{k}$ and $v_{k}$ with
\begin{equation}
v_{k}^{2}=\frac{1}{2}\left(  1-\frac{\varepsilon_{k}-\lambda_{\tau_{k}}}%
{E_{k}}\right)
\end{equation}
and $u_{k}=\sqrt{1-v_{k}^{2}}$. The pairing gaps $\Delta_{k}$ are obtained by
the solution of the gap equation%
\begin{equation}
\Delta_{k}=-\frac{1}{2}\sum\limits_{k^{\prime}}V_{k{\bar{k}},k^{\prime}%
{\bar{k}}^{\prime}}^{pp}\frac{\Delta_{k^{\prime}}}{2E_{k^{\prime}}}
\label{BCS1}%
\end{equation}
in each step of the iteration and the chemical potential $\lambda_{\tau_{k}}$
is fixed via particle number conservation:
\begin{equation}
\sum\limits_{k}v_{k}^{2}=N\text{ \ (or }Z\text{) \ \ \ for neutrons (or
protons)}. \label{BCS3}%
\end{equation}
After the solution of the BCS equations (\ref{BCS1}-\ref{BCS3}) the density
(\ref{rho}) is calculated and used for the solution of the Klein-Gordon
equations (\ref{kg}) determining the RMF potentials for the Dirac-Hartree
Hamiltonian in Eq. (\ref{dhd}) in the next step of the iteration. In the
RMF+BCS approximation the eight components of the
quasiparticle eight-component\textsl{ } Dirac spinor $|\psi_{k}^{\eta}\rangle$
are simply expressed through the usual 4-component spinor
wave functions $\varphi_{k}$:
\begin{align}
U_{k}(x)  &  =u_{k}\varphi_{k}(x)\nonumber\\
V_{k}(x)  &  =(-1)^{l_{k}+j_{k}+m_{k}}v_{k}\varphi_{-k}^{\ast}(x),
\label{dhbcs}%
\end{align}
and we have chosen $u_{k},v_{k}>0\ \forall k$. This simplifies the calculation
of the quasi-particle RPA matrix elements in the next section considerably. We
only have to calculate the matrix elements in particle space using the wave
functions $\varphi_{k}$ and multiply them with the corresponding BCS
occupation factors in Eqs. (\ref{eta}) and (\ref{xi}).

In the present applications of our approach we use a monopole force with
constant matrix elements and a soft pairing window. Details are given below.

\subsection{Solution of the RQRPA equations and calculation of the phonon
vertices}

The RQRPA equations are derived as the small amplitude limit of the
time-dependent Dirac-Hartree-Bogoliubov equations for the generalized density
matrix $\mathcal{R}$ \cite{PRN.03}. For general pairing forces, as for
instance for the finite range Gogny force in the pairing channel \cite{GEL.96}
they can be solved in the canonical basis \cite{RS.80} of the RHB equations,
where the full Hartree-Bogoliubov ground state wave function has BCS form. In
the RMR+BCS case they are solved in the Dirac-Hartree-BCS basis (\ref{dhbcs})
described above. In spherical systems we can use angular momentum coupling of
the 2-quasiparticle states and the reduced form of the RQRPA equation for
angular momentum $J_{\mu}$ is:
\begin{equation}
\bigl(\eta\Omega_{\mu}-E_{k_{1}}-E_{k_{2}}\bigr)\mathcal{R}_{\mu(k_{1}k_{2}%
)}^{\eta}=\sum_{\eta^{\prime}}\sum_{(k_{4})\leq(k_{3})}{\tilde{V}}%
_{(k_{1}k_{4},k_{2}k_{3})}^{J_{\mu},\eta\eta^{\prime}}\mathcal{R}_{\mu
(k_{3}k_{4})}^{\eta^{\prime}}, \label{qrpaj}%
\end{equation}
where the index $\mu$ characterizes the various solutions of the RQRPA
equation, in particular their angular momentum $J_{\mu}$. The notation
$(k_{1}k_{2})$ indicates the fact that the two quasiparticles with the indices
$k_{1}$ and $k_{.2}$ are coupled to angular momentum $J_{\mu}$.
The $\eta\eta^{\prime}$ components of the static residual interaction in the
$ph$-channel read:
\begin{align}
{\tilde{V}}_{(k_{1}k_{4},k_{2}k_{3})}^{J,\eta\eta^{\prime}}  &  =\sum
\limits_{S=0,1}(\delta_{\eta,1}+(-1)^{S}\delta_{\eta,-1})(\delta_{\eta
^{\prime},1}+(-1)^{S}\delta_{\eta^{\prime},-1})\times\nonumber\\
&  \qquad\qquad\qquad\qquad\qquad\times\Bigl[\eta_{(k_{1}k_{2})}^{S}%
\eta_{(k_{3}k_{4})}^{S}{\tilde{v}}_{(k_{1}k_{4},k_{2}k_{3})}^{(ph)JS}%
+\xi_{(k_{1}k_{2})}^{S}\xi_{(k_{3}k_{4})}^{S}{\tilde{v}}_{(k_{1}k_{2}%
,k_{3}k_{4})}^{(pp)J}\Bigr], \label{fintetaeta}%
\end{align}
where ${\tilde{v}}_{(k_{1}k_{4},k_{2}k_{3})}^{(ph)JS}$ and ${\tilde{v}%
}_{(k_{1}k_{2},k_{3}k_{4})}^{(pp)J}$ are the reduced matrix elements of the
$ph$- and $pp$-interaction. We assume that the $pp$-components do not depend
on the total spin, and the $ph$-components carry spin $S=0,1$. The
$ph$-components ${\tilde{v}}_{(14,23)}^{(ph)JS}$ describe the one-boson
exchange (OBE) interaction and could be expressed as follows:
\begin{align}
\tilde{v}_{(k_{1}k_{4},k_{2}k_{3})}^{(ph)JS}  &  =\pm\frac{(4\pi)^{2}}%
{2J+1}\sum\limits_{m\in S}\sum\limits_{L}\,\int\limits_{0}^{\infty}\frac
{q^{2}q^{\prime\ 2}dqdq^{\prime}}{(2\pi)^{6}}\langle(k_{1})\Vert
j_{L}(qr)[\beta\Gamma_{mS}Y_{L}]^{J}\Vert(k_{2})\rangle\times\nonumber\\
&  \qquad\qquad\qquad\qquad\qquad\qquad\qquad\times D_{m}^{S}(q,q^{\prime
})\langle(k_{3})\Vert j_{L}(q^{\prime}r)[\beta\Gamma_{mS}Y_{L}]^{J}\Vert
(k_{4})\rangle, \label{vphj}%
\end{align}
where in the first sum ($m\in S$) the index $m$ \ runs over the various meson
fields carrying spin $S$. The index $S$ in $\Gamma_{mS}$ denotes the spin of
the Pauli matrix entering the vertices $\Gamma_{m}$ in Eq. (\ref{gammas}).
This implies in particular that $S=0$ for the scalar and time-like parts of
the vector mesons and that $S=1$ for the space-like parts of the vector mesons
(current-current interactions).

Representing the $q$-integral in Eq. (\ref{vphj}) by a discrete sum over mesh
points, the matrix elements (\ref{vphj}) are a sum of separable terms. The
non-local meson propagator is a solution of the integral equation:
\begin{equation}
\bm{q}^{2}D_{m}^{S}(\bm{q},\bm{q}^{\prime})+{\displaystyle\int}\frac
{d^{3}q^{\prime\prime}}{(2\pi)^{3}}~M_{m}^{S}(\bm{q-q}^{\prime\prime}%
)D_{m}^{S}(\bm{q}^{\prime\prime},\bm{q}^{\prime})=(2\pi)^{3}\delta
(\bm{q}-\bm{q}^{\prime}),
\end{equation}
where $M_{m}^{S}(\bm{q})$ is the Fourier transform of $U^{\prime\prime}%
(\phi_{m}^{S}(\bm{r}))$ determined by Eq. (\ref{uphi},\ref{NL}):%
\begin{equation}
M_{m}^{S}(\bm{q})={\displaystyle\int}d^{3}re^{-i\bm{qr}}U^{\prime\prime}%
(\phi_{m}^{S}(\bm{r})).
\end{equation}
The quantities $\eta_{(k_{1}k_{2})}^{S},\xi_{(k_{1}k_{2})}^{S}$ in the Eq.
(\ref{fintetaeta}) are the conventional factors \cite{RS.80} which are the
following linear combinations of the occupation numbers:
\begin{align}
\eta_{(k_{1}k_{2})}^{S} &  =\frac{1}{\sqrt{1+\delta_{(k_{1}k_{2})}}%
}\Bigl(u_{k_{1}}v_{k_{2}}+(-1)^{S}v_{k_{1}}u_{k_{2}}\Bigr)\label{eta}\\
\xi_{(k_{1}k_{2})}^{S} &  =\frac{1}{\sqrt{1+\delta_{(k_{1}k_{2})}}%
}\Bigl(u_{k_{1}}u_{k_{2}}-(-1)^{S}v_{k_{1}}v_{k_{2}}\Bigr),\label{xi}%
\end{align}
arising due to symmetrization in the integral part of the Eq. (\ref{qrpaj}),
which enables one to take each $2q$-pair into account only once because of the
symmetry properties of the reduced matrix elements ${\tilde{v}}_{(k_{1}%
k_{4},k_{2}k_{3})}^{(ph)JS}$ and ${\tilde{v}}_{(k_{1}k_{2},k_{3}k_{4}%
)}^{(pp)J}$. For the interaction ${\tilde{v}}^{(pp)}$ in the $pp$-channel we
use a simple monopole-monopole ansatz with the so-called smooth window
\cite{BFH.85}:
\begin{equation}
{\tilde{v}}_{(k_{1}k_{2},k_{3}k_{4})}^{(pp)J}=-\frac{G}{2}\,\delta_{J0}%
\delta_{(k_{1}k_{2})}\delta_{(k_{3}k_{4})}\sqrt{\frac{2j_{k_{1}}%
+1}{1+e^{(\varepsilon_{k_{1}}-w)/d}}}\sqrt{\frac{2j_{k_{3}}+1}%
{1+e^{(\varepsilon_{k_{3}}-w)/d}}},\label{vppj}%
\end{equation}
where $w$ is the value of the pairing window and $d$ is its diffuseness.

The RQRPA transition densities $\mathcal{R}^{\eta}_{\mu(k_{1}k_{2})}$
calculated from the Eq. (\ref{qrpaj}) determine in particular the components
of the amplitudes $\gamma_{\mu(k_{1}k_{2})}^{\eta}$ which couple the phonon
with the quasiparticle states $|\psi^{\eta_{1}}_{k_{1}}\rangle$ and
$|\psi^{\eta_{2}}_{k_{2}}\rangle$ having $\eta_{1} = \eta_{2} = \eta$ i.e.
lying on the same side with respect to the Fermi level
\begin{align}
\gamma_{\mu(k_{1}k_{2})}^{\eta}  & = \sqrt{1 + \delta_{(k_{1}k_{2})}}
\sum_{\eta^{\prime}} \sum_{(k_{4})\leq(k_{3})}\sum\limits_{S=0,1}
(\delta_{\eta,1 } - (-1)^{S}\delta_{\eta,-1}) (\delta_{\eta^{\prime},1 } +
(-1)^{S}\delta_{\eta^{\prime},-1})\nonumber\\
& \qquad\qquad\qquad\qquad\times\Bigl[ \xi^{S}_{(k_{1}k_{2})}\eta^{S}%
_{(k_{3}k_{4})} {\tilde v}^{(ph)J_{\mu}S}_{(k_{1}k_{4},k_{2}k_{3})} - \eta
^{S}_{(k_{1}k_{2})}\xi^{S}_{(k_{3}k_{4})} {\tilde v}^{(pp)J_{\mu}}%
_{(k_{1}k_{2},k_{3}k_{4})}\Bigr] \mathcal{R}^{\eta^{\prime}}_{\mu(k_{3}k_{4}%
)}. \label{gammaj}%
\end{align}


\subsection{The RQTBA correlated propagator and the strength function}

In solving Eq.~(\ref{bse}) for the response function, we use our previous
experience with calculations for nuclei with closed shells
\cite{LRT.07,LRV.07}. Again, we formulate and solve this equation both in the
$2q$-basis of Dirac-Hartree-BCS quasiparticle pairs and in the momemtum-channel
space. In Dirac-Hartree-BCS space its dimension is the number of $2q$-pairs
which satisfy the selection rules for the given multipolarity. In relativistic
nuclear calculations it is always important to take into account the
contribution of the Dirac sea. This can be done, as it is done traditionally,
explicitly, or statically by the renormalization of the static interaction, as
it is proposed in Ref. \cite{SNS}. Nevertheless, for systems with pairing
correlations the total number of $2q$-pairs entering Eq.~(\ref{bse}) increases
considerably not only with the nuclear mass number, but also with the pairing
window. As it was investigated in a series of RRPA calculations
\cite{RMG.01,MWG.02}, the completeness of the $ph$ ($\alpha h$) basis is very
important for calculations of giant resonance characteristics as well as for
current conservation and a proper treatment of symmetries, in particular, the
dipole spurious state originating from the violation of translation symmetry
on the mean field level. On the other hand, the use of a large basis requires
a considerable numerical effort and, therefore, it is reasonable to solve the
Eq.~(\ref{bse}) in a different more appropriate representation.

Our choice is determined by the following properties of the static effective
interaction ${\tilde{V}}$. Its $ph$-component is based on the exchange of
mesons and explicitly contains only direct terms and no exchange terms,
therefore it can be written as a sum of separable interactions (\ref{vphj}),
and in the present work its $pp$-component is also chosen in the separable
form (\ref{vppj}) for convenience.

As in Refs. \cite{LRT.07,LRV.07}, we solve the response equation for a fixed
value of the energy variable $\omega$ in two steps. First, we calculate the
correlated propagator $R^{e}(\omega)$ which describes the propagation under
the influence of the interaction $\Phi(\omega)$ in the time-blocking
approximation without GSC caused by the phonon coupling:
\begin{eqnarray}
R_{(k_{1}k_{4},k_{2}k_{3})}^{(e)J,\eta}(\omega) &=& {\tilde{R}}^{(s)J,\eta}%
_{(k_{1}k_4,k_{2}k_3)}(\omega)\nonumber\\
&+& {\tilde{R}}^{(0)\eta}_{(k_{1}k_{2})}(\omega)
\sum\limits_{(k_{6}\leq k_{5})}
\Bigl[ \Phi_{(k_{1}k_{6},k_{2}%
k_{5})}^{(s)J,\eta}(\omega) - \Phi_{(k_{1}k_{6},k_{2}k_{5})}^{(s)J,\eta}%
(0)\Bigr] R_{(k_{5}k_{4},k_{6}k_{3})}^{(e)J,\eta}(\omega) ,
\label{correlated-propagator}%
\end{eqnarray}
where the symmetrized matrix elements of the mean field propagator
${\tilde R}^{(s)}$ and the two quasiparticles-phonon coupling
amplitude $\Phi^{(s)}$ read:
\begin{eqnarray}
{\tilde{R}}^{(s)J,\eta}_{(k_{1}k_{4},k_{2}k_{3})}(\omega) &=& {\tilde
R}^{(0)\eta}_{(k_1k_2)}(\omega) \bigl(
\delta_{(k_1k_3)}\delta_{(k_2k_4)} + (-)^{J+l_1-l_2+j_1-j_2}
\delta_{(k_1k_4)}\delta_{(k_2k_3)} \bigr),
\\
\Phi^{(s)J,\eta}_{(k_1k_4,k_2k_3)}(\omega) &=& \frac{1}{1 +
\delta_{(k_3k_4)}} \Bigl( \Phi^{J,\eta}_{(k_1k_4,k_2k_3)}(\omega) +
(-)^{J+l_1-l_2+j_1-j_2}\Phi^{J,\eta}_{(k_2k_4,k_1k_3)}(\omega) \Bigr),
\end{eqnarray}
which means that we take into account two kinds of components: one
kind with
only forward ($\eta$ $=1$) $2q$-propagators of the $ph$-type ($\eta_{1}%
=-\eta_{2}$) and another one with only backward propagators ($\eta=-1$), but
do not include mixed ones. In the conventional terminology it means that we
neglect ground state correlations caused by the quasiparticle-phonon coupling.
The reduced matrix elements of the quasiparticle-phonon coupling amplitude
$\Phi_{(k_{1}k_{4},k_{2}k_{3})}^{J,\eta}(\omega)$ read:
\begin{align}
\Phi_{(k_{1}k_{4},k_{2}k_{3})}^{J,\eta}(\omega)  &  =\sum\limits_{\mu
}\biggl[\frac{\delta_{(k_{1}k_{3})}\delta_{{\varkappa}_{k_{4}}{\varkappa
}_{k_{2}}}}{2j_{k_{2}}+1}\sum\limits_{(k_{6})}\frac{\gamma_{\mu(k_{6}k_{2}%
)}^{-\eta}\gamma_{\mu(k_{6}k_{4})}^{-\eta\ast}}{\eta\omega-E_{k_{1}}-E_{k_{6}%
}-\Omega_{\mu}}\nonumber\\
&  \qquad+\frac{\delta_{(k_{2}k_{4})}\delta_{{\varkappa}_{k_{3}}{\varkappa
}_{k_{1}}}}{2j_{k_{1}}+1}\sum\limits_{(k_{5})}\frac{\gamma_{\mu(k_{1}k_{5}%
)}^{\eta}\gamma_{\mu(k_{3}k_{5})}^{\eta\ast}}{\eta\omega-E_{k_{5}}-E_{k_{2}%
}-\Omega_{\mu}}\nonumber\\
& \qquad+(-1)^{J+J_{\mu}}\left\{
\begin{array}
[c]{ccc}%
j_{k_{1}} & j_{k_{2}} & J\\
j_{k_{4}} & j_{k_{3}} & J_{\mu}%
\end{array}
\right\}  \biggl(\frac{(-1)^{j_{k_{3}}-j_{k_{2}}}\gamma_{\mu(k_{1}k_{3}%
)}^{\eta}\gamma_{\mu(k_{2}k_{4})}^{-\eta\ast}}{\eta\omega-E_{k_{3}}-E_{k_{2}%
}-\Omega_{\mu}}\nonumber\\
& \qquad\qquad\qquad\qquad\qquad\qquad\text{ \ \ \ \ \ } +\frac{(-1)^{j_{k_{1}%
}-j_{k_{4}}}\gamma_{\mu(k_{3}k_{1})}^{\eta\ast} \gamma_{\mu(k_{4}k_{2}%
)}^{-\eta}}{\eta\omega-E_{k_{1}}-E_{k_{4}}-\Omega_{\mu}}\biggr)\biggr],
\label{phiphc}%
\end{align}
where $\varkappa_{k}$ denotes the relativistic quantum number set:
$\varkappa_{k}=(2j_{k}+1)(l_{k}-j_{k})$. The reduced matrix elements of the
particle-phonon coupling amplitude $\gamma_{\mu(k_{1}k_{2})}^{\eta}$ are
calculated from the Eq. (\ref{gammaj}). The index $\mu=\{J_{\mu},n_{\mu}\}$
denotes the set of phonon quantum numbers which are its angular momentum
$J_{\mu}$ and the number of the solution $n_{\mu}$ of the Eq. (\ref{qrpaj}).
The quantity $\Omega_{\mu}$ is the corresponding energy. The fact, that the
r.h.s. of Eq. (\ref{phiphc}) depends only on the same $\eta$-values as the
l.h.s. and does not contain any mixing of different $\eta$-values implies that
no GSC are contained in the intermediate $2q\otimes phonon$ propagators.

The Eq. (\ref{correlated-propagator}) is too expensive numerically to be
solved in the full Dirac-Hartree-BCS basis. However, due to the pole structure
of the $\Phi$-amplitude it is naturally to suggest that quasiparticle-phonon
coupling effects are not important quantitatively far from the Fermi surface.
In the present work, for numerical calculations an energy window $E_{win}$ was
implemented around the Fermi surface with respect to pure two-quasiparticle
energies $E_{2q}$ so that the summation in the
Eq.~(\ref{correlated-propagator}) is performed only among the $2q$-pairs with
$E_{2q}\leq E_{win}$. Consequently, the correlated propagator differs from the
mean field propagator only within this window. This approximation has been
checked in the Ref. \cite{LRT.07} in the calculations for nuclei with closed
shells by direct calculations with different values of this energy window, and
it has been found that this window should include just the investigated energy
region. Beyond the energy window we do not obtain additional poles caused by
$2q\otimes phonon$ configurations, but only the renormalized QRPA spectrum. It
is important to emphasize that many 2$q$- and $\alpha q$-configurations outside
of the window are taken into account on the RQRPA level that is necessary in
order to obtain the reasonable centroid positions of giant resonances as well
as to find the dipole spurious state close to zero energy. By its physical
meaning, the Eq. (\ref{correlated-propagator}) contains all effects of the
quasiparticle-phonon coupling and all the singularities of the integral part
of the initial BSE.

In the second step, we have to solve the remaining equation for the
full response function $R(\omega)$:
\begin{align}
R_{(k_{1}k_{4},k_{2}k_{3})}^{J,\eta\eta^{\prime}}(\omega) &=
R_{(k_{1}k_{4},k_{2}k_{3})}^{(e)J,\eta}(\omega)\delta_{\eta\eta^{\prime}}
\nonumber\\
&+ \sum\limits_{(k_{6}\leq k_{5})} \sum\limits_{(k_{8}\leq
k_{7})\eta^{\prime\prime}}R_{(k_{1}k_{6},k_{2}k_{5})}^{(e)J,\eta}(\omega)
{\tilde V}^{J,\eta\eta^{\prime\prime}}_{(k_{5}k_{8},k_{7}k_{6})} R_{(k_{7}%
k_{4},k_{8}k_{3})}^{J,\eta^{\prime\prime}\eta^{\prime}}(\omega).
\label{responsej}%
\end{align}
In contrast to the Eq.~(\ref{correlated-propagator}), this equation
contains only the static effective interaction $\tilde V$ from the
Eq. (\ref{fintetaeta}).

Since both the one-boson exchange interaction and the pairing
interaction are separable in momentum space, we can use this
advantage and formulate the response equation in the
momentum-channel representation. Let us introduce the following
generalized channel index $c_{\chi} = \{  q,m,L,S \}$ for $\chi =
\left(ph\right)$ and $c_{\chi} = S$ for $\chi = \left( pp \right)$.
For $\chi = \left( ph \right)$ it includes the momentum $q$
transferred in the exchange process of the corresponding meson
labeled by the index $m$. The index $\chi$ distinguishes $ph$- and
$pp$-channel components of the static interaction, L is the angular
momentum, and the index S=0,1 has its usual meaning of the total
spin carried through the certain channel. In this way, we apply the
following ansatz for the $\eta$-components of the static effective
interaction ${\tilde V}$
\begin{equation}
{\tilde V}^{(J)\eta\eta^{\prime}}_{(k_{1}k_{4},k_{2}k_{3})}=\sum
\limits_{cc^{\prime}}Q_{(k_{1}k_{2})}^{(c)J,\eta} d_{cc^{\prime}}Q_{(k_{3}%
k_{4})}^{(c^{\prime})J,\eta^{\prime}\ast} ,
\end{equation}
where we omit the index $\chi$ for simplicity. For the channels with
$\chi = (ph)$:
\begin{align}
Q_{(k_{1}k_{2})}^{(c)J,\eta}  &  = \frac{\delta_{\eta,1 } +
(-1)^{S}\delta_{\eta
,-1}}{\sqrt{1 + \delta_{(k_1k_2)}}}\eta^{S}_{(k_{1}k_{2})} \langle(k_{1})\Vert j_{L}(qr)[\beta\Gamma_{mS}%
Y_{L}]^{J}\Vert(k_{2})\rangle\\
d_{cc^{\prime}}  &  = \pm\frac{1}{2J+1}\frac{D_{m}^{S}(q,q^{\prime})}{(2\pi)^{6}}%
\delta_{LL^{\prime}}\delta_{SS^{\prime}}\delta_{mm^{\prime}}%
\end{align}
and the summation over $c,c^{\prime}$ implies integration over
$d^{3}q, d^{3}q^{\prime}$. For the channels with $\chi= (pp)$ we
have:
\begin{align}
Q_{(k_{1}k_{2})}^{(c)J,\eta}  &  =
\delta_{J0}\delta_{(k_1k_2)}\frac{\delta_{\eta,1 } +
(-1)^{S}\delta_{\eta,-1}}{\sqrt{1 +
\delta_{(k_1k_2)}}}\xi^{S}_{(k_{1}k_{2})} \sqrt
{\frac{2j_{k_{1}}+1}{1 + e^{(\varepsilon_{k_{1}} - w)/d}}}\\
d_{cc^{\prime}}  &  = -\frac{G}{2}\,\delta_{cc^{\prime}} .
\end{align}

Then, we can use the well known techniques of the response formalism
with separable interactions (see, for instance, Ref.~\cite{RS.80}).
We define the exact response function and the correlated propagator
in the generalized
momentum-channel space as follows:%
\begin{align}
R_{cc^{\prime}}^{J}(\omega) &  =\sum\limits_{(k_{2}\leq
k_{1})\eta}\sum
\limits_{(k_{4}\leq k_{3})\eta^{\prime}}Q_{(k_{1}k_{2})}^{(c)J,\eta\ast}R_{(k_{1}%
k_{4},k_{2}k_{3})}^{J,\eta\eta^{\prime}}(\omega)Q_{(k_{3}k_{4})}^{(c^{\prime})J,\eta^{\prime}}\\
R_{cc^{\prime}}^{(e)J}(\omega) &  =\sum\limits_{(k_{2}\leq k_{1})}%
\sum\limits_{(k_{4}\leq k_{3})\eta}Q_{(k_{1}k_{2})}^{(c)J,\eta\ast}%
R_{(k_{1}k_{4},k_{2}k_{3})}^{(e)J,\eta}(\omega)Q_{(k_{3}k_{4})}^{(c^{\prime
})J,\eta}.\label{rcc}%
\end{align}
In this representation Eq.~(\ref{responsej}) reads:%
\begin{equation}
R_{cc^{\prime}}^{{}} = R_{cc^{\prime}}^{e} +
(R^{e}dR)_{cc^{\prime}}.
\end{equation}
This equation is solved by matrix inversion%
\begin{equation}
R=\Bigl(1-R^{e}d\Bigr)^{-1}R^{e}.\label{bse4}%
\end{equation}
To compute the nuclear response in the certain external field, we
need a convolution of the exact response function with the external
field operator $P$ which can be suggested as an additional channel
$c=p$, $p=\{z,\chi\}$, where the index $z$ contains possible
additional dependences of the external field which we do not
concretize here:
\begin{equation}
P_{(k_{1}k_{2})}^{(p)J,\eta}=\sum\limits_{LS}
\frac{\delta_{\eta,1}+(-1)^{S}\delta_{\eta,-1}}{\sqrt{1 +
\delta_{(k_1k_2)}}} \eta^{S}_{(k_1k_2)}\langle(k_{1})\parallel
P^{(p)J}_{LS}\parallel(k_{2})\rangle.
\end{equation}
Making use of this definition, we can determine the polarizability
as:
\begin{equation}
\Pi^{J}(\omega)=R_{pp}^{J}(\omega)=R_{pp}^{(e)J}(\omega)+\sum
\limits_{cc^{\prime}}R_{pc}^{(e)J}(\omega)d_{cc^{\prime}}R_{c^{\prime}p}%
^{J}(\omega),\label{pol}%
\end{equation}
where the quantities $R_{pc}^{(e)J}(\omega)$,
$R_{pp}^{(e)J}(\omega)$ can be found as follows:
\begin{eqnarray}
R_{pc}^{(e)J}(\omega)=\sum\limits_{(k_{2}\leq
k_{1})}\sum\limits_{(k_{4}\leq
k_{3})\eta}P_{(k_{1}k_{2})}^{(p)J,\eta\ast}R_{(k_{1}k_{4}%
,k_{2}k_{3})}^{(e)J,\eta}(\omega)Q_{(k_{3}k_{4})}^{(c)J,\eta}\nonumber\\
R_{pp}^{(e)J}(\omega)=\sum\limits_{(k_{2}\leq
k_{1})}\sum\limits_{(k_{4}\leq
k_{3})\eta}P_{(k_{1}k_{2})}^{(p)J,\eta\ast}R_{(k_{1}k_{4}%
,k_{2}k_{3})}^{(e)J,\eta}(\omega)P_{(k_{3}k_{4})}^{(p)J,\eta},
\end{eqnarray}
and the quantity $R_{cp}^{J}(\omega)$, which has a meaning of the
density matrix variation in the external field $P$, obeys the
equation:
\begin{equation}
R_{cp}^{J}(\omega)=R_{cp}^{(e)J}(\omega)+\sum\limits_{c^{\prime}%
c^{\prime\prime}}R_{cc^{\prime}}^{(e)J}(\omega)d_{c^{\prime}c^{\prime\prime}%
}R_{c^{\prime\prime}p}^{J}(\omega). \label{rcp}%
\end{equation}
To describe the observed spectrum of the excited nucleus in a weak
external field $P$, as for instance a dipole field, one needs to
calculate the strength function:
\begin{equation}
S^{J}(E)=-\frac{1}{\pi}\lim\limits_{\Delta\rightarrow+0}Im\ \Pi^{J}%
(E+i\Delta),\label{strfj}%
\end{equation}
expressed through the polarizability $\Pi^{J}(\omega)$ defined by
Eq. (\ref{pol}).

Obviously, the dimension of vectors and matrices entering
Eq.~(\ref{rcp}) is determined by the number of mesh-points in
$q$-space and the number of $m,L,S$-channels. In particular, it does
not depend considerably on the total dimension of $2q$- and 
$\alpha q$-subspaces and on the mass number of the
nucleus. As we have realized in the calculations of Refs. \cite{LRT.07,LRV.07}%
, the advantage of the momentum-channel representation appears at some medium
values of the nuclear mass number, where the total dimension of $ph$- and
$\alpha h$-subspaces, which is exactly the dimension of arrays in the
Eq.~(\ref{respdir}) written in the coupled form, become comparable with the
dimension of matrices entering Eq.~(\ref{rcp}). In the present approach, due
to the pairing correlations, this mass region shifts towards lower masses. The
solution in the momentum-channel space is even more helpful when we include
pairing correlations, since the number of states within the pairing window
increases with more than a factor two as compared to the case without pairing.
For heavy nuclei the dimension of the two-quasiparticle DHB basis increases
considerably and, therefore, for heavy nuclei the solution of the response
equations in momentum space is recommendable.

Notice, that the pairing correlations cause also an additional numerical
effort in Eq. (\ref{correlated-propagator}). It is solved within the subspace
of $2q$-configurations confined by the $E_{win}$ which, in the realistic
calculations, surrounds the pairing window and, therefore, contains
considerably more configurations as compared to the case of no pairing.

\section{Computational details, results and discussion}

\label{details}

\subsection{Numerical details}

For this first application we have chosen two chains of spherical even-even
semi-magic nuclei: one chain with $Z=50$ and another one with $N=50$. We have
calculated the isovector dipole spectrum in the giant dipole resonance region
and in the low-lying energy region in the two approximations: RQRPA and RQTBA
for the quasiparticle-vibration coupling. All the results presented below have
been obtained with making use of the NL3 parameter set \cite{NL3} for the
covariant density functional (\ref{ERMF}).

In the present work, pairing correlations were treated in the BCS
approximation where the single quasiparticle wave functions diagonalize the
single-nucleon density matrix ${\rho}$. As pairing interaction $V^{pp}$ we use
the simple monopole-monopole form (\ref{vppj}) within the smoothed energy
window with the parameters $w=20$ MeV, $d=1$ MeV. The parameter $G$ was chosen
in such a way that the resulting gap at the Fermi surface reproduces the
empirical gap expressed by the well known three-point formula:
\begin{equation}
\Delta_{N_{\tau}}^{(3)}=-\frac{(-1)^{N_{\tau}}}{2}[B(N_{\tau}-1)+B(N_{\tau
}+1)-2B(N_{\tau})],
\end{equation}
where $B(N_{\tau})$ is the experimentally known binding energy of the nucleus
with $N_{\tau}$ nucleons in the subsystem with pairing correlations (neutrons
or protons). The RMF plus BCS equations are solved by expanding the nucleon
spinors in a spherical harmonic oscillator basis \cite{GRT.90}. In the present
calculation we have used the basis of $20$ oscillator shells.

In solving the RQRPA Eq. (\ref{qrpaj}) we have used the method proposed in
Ref. \cite{Pap.07} for a reduction of the eigenvalue problem by the
generalized Cholesky decomposition. In the RQRPA as well as RQTBA calculations
both Fermi and Dirac subspaces were truncated at energies far away from the
Fermi surface: in the present work as well as in the
Refs.~\cite{LRT.07,LRV.07} we fix the limits $E_{2q}<100$ MeV and $E_{\alpha
q}>-1800$ MeV with respect to the positive continuum (so far from the Fermi
surface there are no pairing effects, therefore we have there pure particles
and holes). A small artificial width was introduced as an imaginary part of
the energy variable $\hbar\omega$ to have a smooth envelope of the calculated
curves. In the calculations for tin isotopes we took 200 keV smearing for the
spectrum in the wide energy region 0-30 MeV and 20 keV for the low-lying
portion of the same spectrum below 10 MeV to distinguish its fine structure.
For the N=50 isotopes we have used the smearing 400 keV, assuming the more
pronounced contribution of the single-particle continuum in the GDR region,
and 10 keV for the low-lying strength.

The energies and amplitudes of the most collective phonon modes with spin and
parity 2$^{+}$, 3$^{-}$, 4$^{+}$, 5$^{-}$, 6$^{+}$ have been calculated with
the same restrictions and selected using the same criterion as in the
Ref.~\cite{LRT.07,LRV.07} and in many other non-relativistic investigations in
this context. Only the phonons with energies below the neutron separation
energy for the examined tin isotopes and below 10 MeV -- for the $N=50$ nuclei
enter the phonon space since the contributions of the higher-lying modes are
supposed to be small. Our previous experience within the non-relativistic
approach of Ref. \cite{LT.07} without the restriction of the phonon space by
the energy have shown that the inclusion of the high-lying modes into the
phonon space cause the change of the mean energies and widths of the
resonances comparable with the smearing parameter (imaginary part of the
energy variable) used in the calculations, because the physical sense of this
parameter is to emulate contributions of remaining configurations which are
not taken into account explicitly.

As a test of numerical correctness of our codes, the response equation has
been solved both in the DHBCS basis and in momentum-channel space and identical
results have been obtained. Since the quasiparticle-phonon coupling amplitude
(\ref{phiphc}) has a pole structure, its contributions to the final result for
the strength function decrease considerably when we go away from the Fermi
surface. Therefore, this coupling has been taken into account only within the
$2q$-energy window $E_{2q}\leq$ 25 MeV around the Fermi surface. This
restriction means that above this energy we have no poles induced by the
complex configurations, and obtain the pure RQRPA poles, but with larger
strength which comes from the integral contribution of the lower-lying energy
spectrum. It has been checked that a further increase of this window does not
influence considerably the strength functions at energies below the value of
this window.

Although a large number of configurations of the $2q\otimes phonon$ type are
taken into account explicitly in our approach, nevertheless we stay in the
same two-quasiparticle space as in the RQRPA, therefore the problem of
completeness of the phonon basis does not arise and, therefore, the phonon
subspace and the subspace of the $2q\otimes phonon$ states can be truncated in
the above mentioned way. Another essential point is, that on all three stages
of our calculations the same relativistic nucleon-nucleon static interaction
$\tilde{V}$ has been employed. The vertices (\ref{gammaj}) entering the QPC
energy-dependent interaction are calculated with the same force. Therefore no
further parameters are needed, and our calculation scheme is fully consistent.

The subtraction procedure developed in the Ref.~\cite{Tse.07} for
self-consistent schemes has been incorporated in our approach. As it was
mentioned above, this procedure removes the static contribution of the
quasiparticle-phonon coupling from the static interaction in the $ph$-channel.
Therefore, the QPC interaction takes into account only the additional energy
dependence introduced by the dynamics of the system. It has been found in the
present calculations as well as in the calculations of the Ref.~\cite{LT.07}
that within the relatively large energy interval (0 - 30 MeV) the subtraction
procedure provides a rather small increase of the mean energy of the giant
dipole resonance (about 0.7 MeV for tin region) and gives rise to the change
by a few percents in the sum rule. This procedure restores the response at
zero energy and, therefore, it does not disturb the symmetry properties of the
RQRPA calculations. The zero energy modes connected with the spontaneous
symmetry breaking in the mean field solutions, as, for instance, the
translational mode in the dipole case, remain at exactly the same positions
after the inclusion of the quasiparticle-vibration coupling. In practice,
however, because of the limited number of oscillator shells in our
calculations this state is found already in the RQRPA without the QPC at a few
hundreds keV above zero. In cases, where the results depend strongly on a
proper separation of this spurious state, as, for instance, for investigations
of the pygmy dipole resonance in neutron rich systems, we have to include a
large number of the $2q$-configurations in the RQRPA solution to avoid mixing of
the spurious state with the low-lying physical states.

\subsection{Isovector dipole strength distribution in semi-magic nuclei:
pygmy and giant resonances}

\begin{figure}[ptb]
\begin{center}
\includegraphics*[scale=1.75]{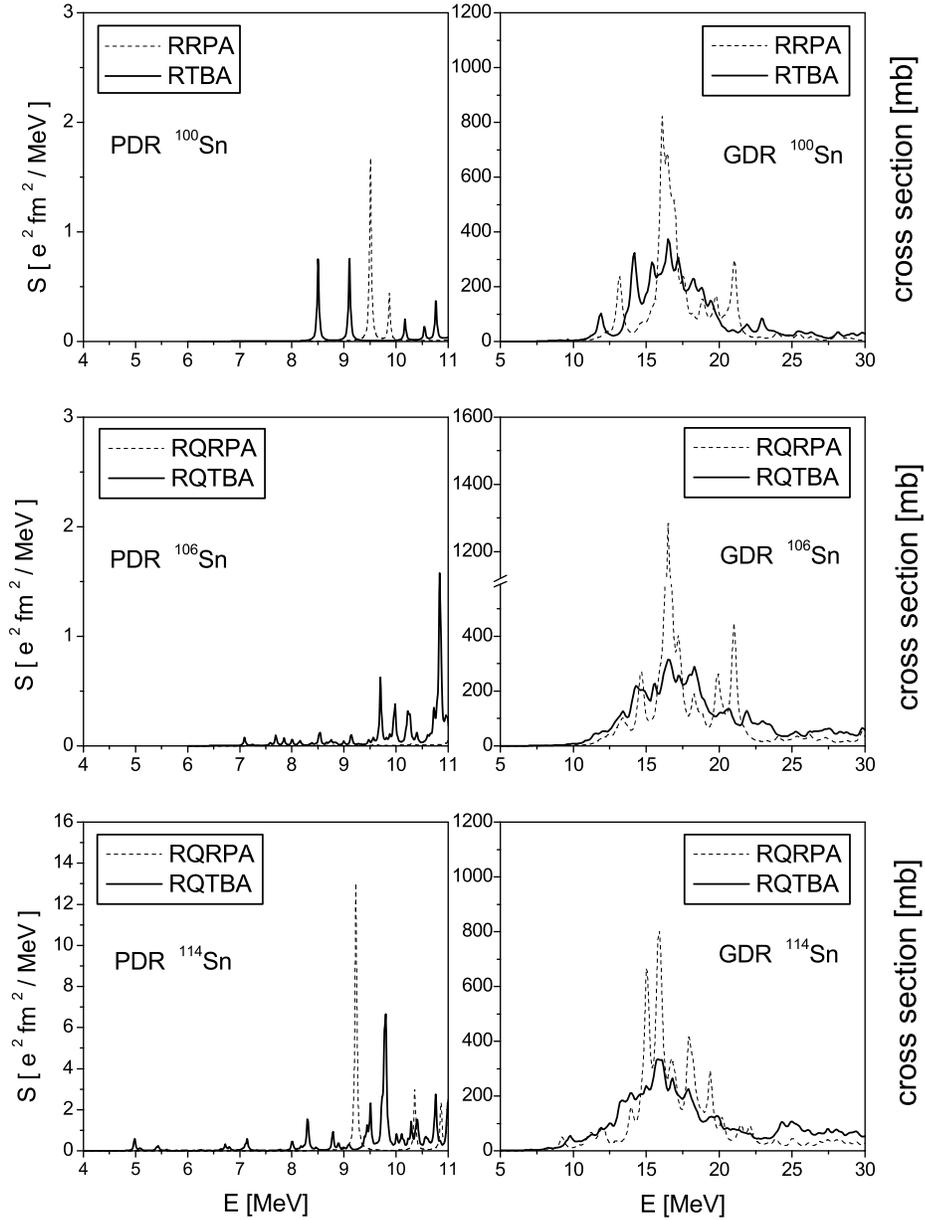}
\end{center}
\caption{The calculated dipole spectra for the light tin isotopes. Right
panels: photo absorption cross sections computed with the artificial width 200
keV. Left panels: the low-lying portions of the corresponding spectra in terms
of the strength function, calculated with 20 keV smearing. Calculations within
the RQRPA are shown by the dashed curves, and the RQTBA - by the solid curves.
\cite{exfor}. }%
\label{f1}%
\end{figure}

\begin{figure}[ptb]
\begin{center}
\includegraphics*[scale=1.75]{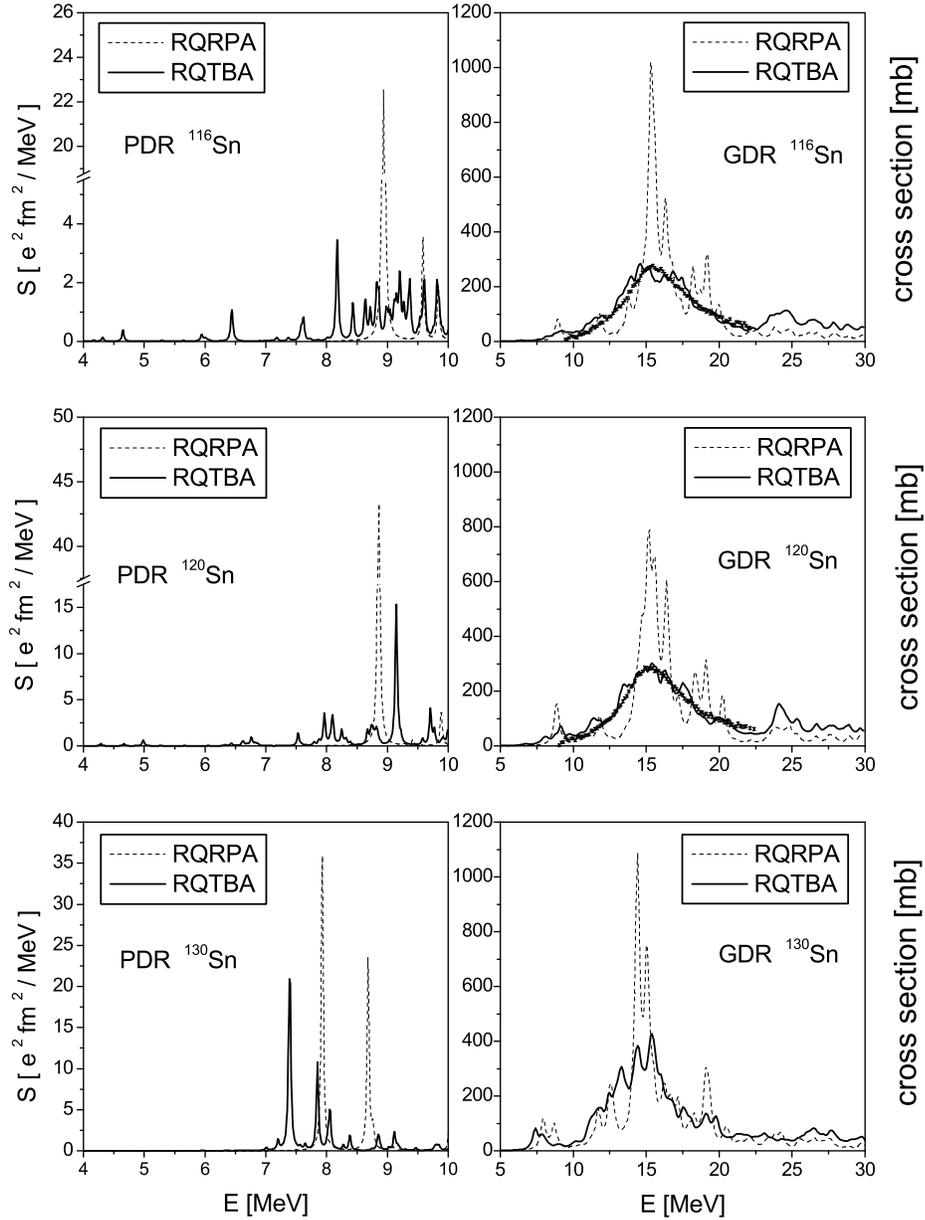}
\end{center}
\caption{The same as in Fig. \ref{f1}, but for heavier tin isotopes, compared
to data of Ref. \cite{exfor} for $^{116,120}$Sn. }%
\label{f2}%
\end{figure}

\begin{figure}[ptb]
\begin{center}
\includegraphics*[scale=1.75]{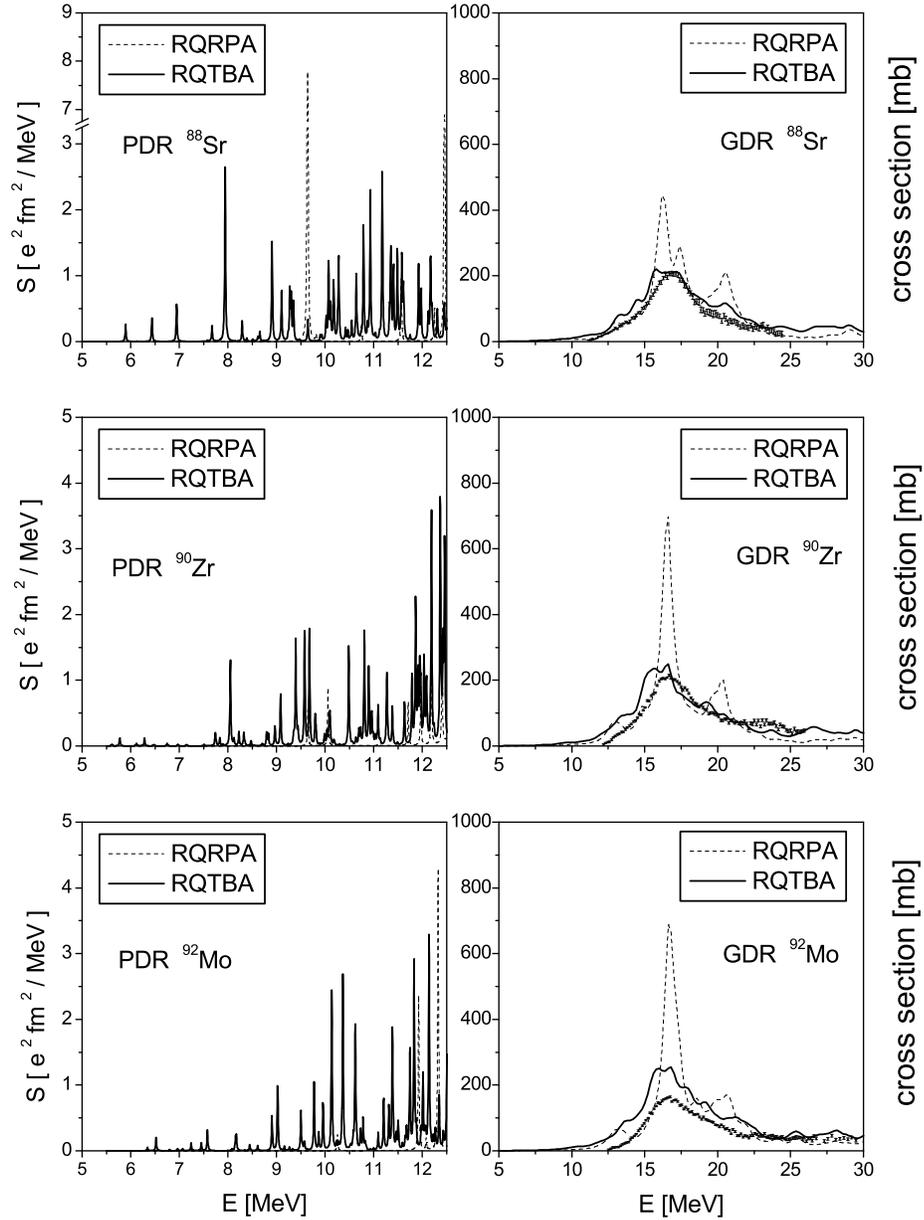}
\end{center}
\caption{The same as in Fig. \ref{f1}, but for $N=50$ isotopes, compared to
data from Ref. \cite{exfor}. }%
\label{f3}%
\end{figure}

In Figs. \ref{f1} and \ref{f2} the calculated dipole spectra for the tin
isotopes $^{100}$Sn, $^{106}$Sn, $^{114}$Sn and $^{116}$Sn, $^{120}$Sn,
$^{130}$Sn, respectively, are given. The right panels of the figures show the
photo absorption cross section
\begin{equation}
\sigma_{E1}(E)={\frac{{16\pi^{3}e^{2}}}{{9\hbar c}}}E~S_{E1}(E),
\end{equation}
which is determined by the dipole strength function $S_{E1}$,
calculated with the usual isovector dipole operator. The left panels
show the low-lying parts of the corresponding spectrum in terms of
the strength function, calculated with the small imaginary part for
the energy variable, in order to see the fine structure of the
spectrum and sometimes individual levels in this region. Fig.
\ref{f3} represents the analogous results for the three $N=50$
nuclei: $^{88}$Sr, $^{90}$Zr and $^{92}$Mo. Calculations within the
RQRPA are shown by the dashed curves, and the RQTBA - by the solid
curves. Experimental data are taken from the EXFOR database
\cite{exfor}.

These three figures clearly demonstrate how the two-quasiparticle states,
which are responsible for the spectrum of the RQRPA excitations, are
fragmented through the coupling to the collective vibrational states. The
effect of the particle-vibration coupling on the low-lying dipole strength
below and around the neutron threshold within the presented approach is shown
in the left panels of the Figs. \ref{f1}-\ref{f3}. Our calculations for the
tin chain give us an example how the low-lying strength develops with the
increase of the neutron excess. In the doubly-magic $^{100}$Sn two first
relatively weak RRPA peaks appear between 9 and 10 MeV. Quasiparticle-phonon
coupling redistributes these structures and shifts them about one MeV lower.
In the $^{106}$Sn due to the pairing correlations in the neutron system the
whole RQRPA picture is shifted towards higher energies, and there is
practically no strength below 10 MeV. In the corresponding figure we find only
the strength caused by the fragmentation of the higher-lying RQRPA peaks above
11 MeV. In the $^{114}$Sn the neutron excess becomes enough to form the
pronounced pygmy mode situated in the RQRPA at about 9.2 MeV and spread over
many states of the $2q\otimes phonon$ nature beginning from 5 MeV. Fig.
\ref{f2} shows how this tendency develops in the more neutron-rich nuclei:
more strength is split to this region and this strength goes to lower energies.

The Lorentz fit parameters for the calculated GDR in the energy intervals:
(10-22.5) MeV for the tin chain and (10-25) MeV for the $N=50$ chain are
displayed in Table~\ref{tab1} and they are compared with the corresponding
data of Refs.~\cite{ripl,Adr.05}. In our work the Lorentz fit is performed in
such a way that the obtained Lorentzian has the same momenta of -2,-1 and zero
orders as our microscopical strength function. This method works well if the
model strength function is rather close to the Lorentz shape. From the
Table~\ref{tab1} we notice that the inclusion of the particle-phonon coupling
in the RQTBA calculation induces a pronounced fragmentation of the photo
absorption cross sections, and brings the mean energies and widths of the GDR
in much better agreement with the data, for all the examined nuclei.

The contribution of the low-lying strength below 10 MeV to the dipole spectrum
is quantified in Table~\ref{tab2}. For the each nucleus, we have calculated
the following quantities: the non-energy weighted sum $\sum B(E1)\uparrow$,
which is obtained by direct integration of the strength, and the
energy-weighted quantity $\sum EB(E1)\uparrow$, which is an integral of the
cross section expressed in the percentage of the classical Thomas-Reiche-Kuhn
sum rule. Both quantities have been calculated with RQRPA and RQTBA to
emphasize the effect of the quasiparticle-phonon coupling for the two energy
intervals: (0-10) MeV and (0-8) MeV. The choice of the intervals is determined
by the fact that, from one hand, in our approach we associate the pygmy modes
with the dipole strength which originates from the first pronounced RQRPA peaks of
the isoscalar nature, and these peaks are situated in the tin isotopes just
below 10 MeV. From the other hand, the measurements of the low-lying strength
excited in these nuclei in the real-photon scattering experiments
\cite{Gov.98,KSS.04} are restricted by the energy around 8 MeV because at
higher energies the sensitivity of these experiments decreases considerably.
Therefore, we have included the strength, calculated in the both energy
intervals, into the Table \ref{tab2}.

The integral contribution of the low-energy portions calculated
within the RQTBA agrees very well with the available data that can
be seen from Table \ref{tab2}: the inclusion of the coupling to
phonons noticeably improves the description. Moreover, below 8 MeV
in the most of the examined nuclei we observe that the
quasiparticle-phonon coupling is the only mechanism which brings the
strength to this region where the pure RQRPA has no solutions at
all. We have found also general agreement of our results for $^{116,130}$Sn
and $^{88}$Sr with the relatively recent studies of the low-lying
dipole strength in Refs. \cite{TLS.04,TL.07} in the Quasiparticle
Phonon Model (QPM) \cite{Sol.92}, although our analysis of transition
densities leads to somewhat different conclusions (this analysis will
be considered in a special publication). In the QPM the model space included up
to three-phonon configurations built from a basis of QRPA states,
calculated with the separable multipole-multipole residual
interactions with adjustable parameters, that could be a possible source
of the above mentioned differences. In these works a clear dependence of the
PDR strength and centroid energies on the neutron-skin thickness was
demonstrated.

Although the integral strength is described rather good within our approach,
the level densities of the obtained low-lying spectra seem to be
underestimated as compared to the experimental works. In the other words, in
our approach the effect of fragmentation of the RQRPA excitations due to
coupling to phonons is not enough. To obtain a more realistic effect, we
could, obviously, use the experience of the previous calculations within the
non-relativistic models. In the Refs. \cite{Gov.98,KSS.04}, the calculations
within the QPM model \cite{Sol.92} with taking into account one-, two- and, in
the Ref. \cite{Gov.98}, also three-phonon configurations lead to the better
description of the PDR fine structure, although some parameters of the
residual interaction were fitted. We could include at least more vibrational
modes into our phonon subspace, that will not even require any modification of
the model. Another way to enrich the spectrum is to take into account ground
state correlations of the singular type, according to Ref. \cite{Tse.07}.

\begin{table}[ptb]
\caption{Characteristics of the isovector dipole spectrum for the
examined $N=50$ and $Z=50$ nuclei: mean energies $\langle E
\rangle$, widths $\Gamma$ and EWSR values calculated with the RQRPA
and with the RQRPA extended by the particle-phonon coupling (RQTBA),
compared to data. The values of $\langle E \rangle$ and $\Gamma$
have been obtained by a Lorentz fit of the computed strength
functions within the interval $[S_n,3S_n]$ where $S_n$ is the
neutron separation energy.}%
\label{tab1}
\begin{center}
\tabcolsep=1.75em \renewcommand{\arraystretch}{0.8}%
\begin{tabular}
[c]{ccccc}\hline\hline
&  & $\langle E \rangle$ & $\Gamma$ & EWSR\\
&  & (MeV) & (MeV) & (\%)\\\hline
& RQRPA & 17.36 & 3.46 & 125\\
$^{88}$Sr & RQTBA & 17.08 & 5.10 & 112\\\hline
& RQRPA & 17.03 & 3.15 & 124\\
$^{90}$Zr & RQTBA & 16.72 & 4.77 & 110\\
& Exp. \cite{ripl} & 16.74 & 4.16 & \\\hline
& RQRPA & 17.45 & 3.09 & 128\\
$^{92}$Mo & RQTBA & 17.13 & 4.72 & 113\\
& Exp. \cite{ripl} & 16.82 & 4.14 & \\\hline
& RRPA & 16.88 & 2.99 & 117\\
$^{100}$Sn & RTBA & 16.39 & 3.43 & 106\\\hline
& RQRPA & 17.17 & 3.07 & 127\\
$^{106}$Sn & RQTBA & 16.53 & 4.89 & 111\\\hline
& RQRPA & 16.35 & 3.67 & 126\\
$^{114}$Sn & RQTBA & 15.80 & 5.42 & 106\\\hline
& RQRPA & 15.95 & 3.11 & 121\\
$^{116}$Sn & RQTBA & 15.35 & 5.17 & 102\\
& Exp. \cite{ripl} & 15.56 & 5.08 & \\\hline
& RQRPA & 15.88 & 3.05 & 121\\
$^{120}$Sn & RQTBA & 15.31 & 5.33 & 104\\
& Exp. \cite{ripl} & 15.37 & 5.10 & \\\hline
& RQRPA & 15.13 & 3.49 & 115\\
$^{130}$Sn & RQTBA & 14.66 & 4.74 & 108\\
& Exp. \cite{Adr.05} & 15.9(5) & 4.8(1.7) & \\\hline\hline
\end{tabular}
\end{center}
\end{table}

\begin{table}[ptb]
\caption{Integral characteristics of the isovector dipole spectrum for the
examined $N=50$ and $Z=50$ nuclei: the integrated strength of the low-lying part
below 10 MeV, calculated with the RQRPA and with the RQRPA extended by the
particle-phonon coupling (RQTBA), compared to the available data.}%
\label{tab2}
\begin{center}
\tabcolsep=1.2em \renewcommand{\arraystretch}{1.0}%
\begin{tabular}
[c]{cccccc}\hline\hline
&  & \multicolumn{2}{c}{(0 - 10) MeV} &
\multicolumn{2}{c}{(0 - 8) MeV}\\
&  & $\sum B(E1)\uparrow$ & $\sum EB(E1)\uparrow$ & $\sum B(E1)\uparrow$ &
$\sum EB(E1)\uparrow$\\
&  & $(e^{2}fm^{2})$ & (\%) & $(e^{2}fm^{2})$ & (\%)\\\hline
& RQRPA & 0.26 & 0.80 & 0.00 & 0.00\\
$^{88}$Sr & RQTBA & 0.32 & 0.80 & 0.13 & 0.30\\
& Exp. \cite{KSS.04} &  &  & 0.141(15) & 0.38(3)\\\hline
& RQRPA & 0.02 & 0.07 & 0.00 & 0.00\\
$^{90}$Zr & RQTBA & 0.34 & 0.90 & 0.07 & 0.15\\\hline
& RQRPA & 0.0049 & 0.01 & 0.00 & 0.00\\
$^{92}$Mo & RQTBA & 0.20 & 0.50 & 0.03 & 0.06\\\hline
& RRPA & 0.14 & 0.30 & 0.00 & 0.00\\
$^{100}$Sn & RTBA & 0.11 & 0.30 & 0.00 & 0.00\\\hline
& RQRPA & 0.01 & 0.03 & 0.00 & 0.00\\
$^{106}$Sn & RQTBA & 0.14 & 0.30 & 0.02 & 0.04\\\hline
& RQRPA & 0.84 & 2.00 & 0.00 & 0.00\\
$^{114}$Sn & RQTBA & 1.38 & 3.00 & 0.20 & 0.30\\\hline
& RQRPA & 1.78 & 4.00 & 0.00 & 0.00\\
$^{116}$Sn & RQTBA & 1.94 & 4.00 & 0.27 & 0.40\\
& Exp. \cite{Gov.98} &  &  & 0.204(25) & \\\hline
& RQRPA & 3.04 & 6.00 & 0.00 & 0.00\\
$^{120}$Sn & RQTBA & 3.08 & 6.00 & 0.62 & 1.00\\\hline
& RQRPA & 4.04 & 7.00 & 2.09 & 4.00\\
$^{130}$Sn & RQTBA & 3.44 & 6.00 & 2.37 & 4.00\\
& Exp. \cite{Adr.05} & 3.2 & 7(3) &  & \\\hline\hline
\end{tabular}
\end{center}
\end{table}

The fragmentation of the resonances, induced by the quasiparticle-phonon
coupling, is a very well known result which has been obtained long ago
\cite{SSV.77,BBBD.79,BB.81} (see also relatively recent calculations of
electric dipole excitations in open-shell nuclei including QPC within the framework of
non-relativistic approaches based on the Skyrme energy functional
\cite{CB.01,SBC.04} as well as on the simple semi-phenomenological scheme
including the single-particle continuum \cite{LT.07}).
Actually, one finds more or less a similar level of agreement
between the available experimental data and the theoretical
predictions of these approaches. We notice, however, that, in
general, our self-consistent relativistic approach reproduces the
shapes and often the mean energies of giant dipole resonances better
than the other above mentioned approaches, that could be attributed
to the more realistic form of the meson-exchange force and to the
fully consistent calculation scheme.

From Figs. \ref{f2}, \ref{f3} one can see that the envelopes of the
calculated GDR in $^{116,120}$Sn between 10 and 22 MeV and in
$^{90}$Zr, $^{88}$Sr between 12 and 25 MeV are rather close to the
experimental cross sections. The deviations from the smooth Lorentz
shape observed in experiments could be attributed to some minor
drawbacks of our approach and calculation scheme: neglecting of the
more complicated, than the $2q\otimes phonon$, configurations by the
time blocking, discretized continuum, restriction of the phonon
subspace by the only low-lying modes, and, at last, too simple model
for the pairing force. Nevertheless, we find that the agreement with
the data for the GDR cross sections in these nuclei is very good,
especially taking into account the fact, that our approach is fully
consistent and contains no any fit additionally to the fit of the
RMF energy functional parameters NL3 which are fixed in the very
beginning and used for the entire nuclear chart. Therefore, we
conclude that the main mechanisms which are responsible for the
damping of the GDR are taken into account correctly and
consistently.

\section{Outlook and conclusions}

\label{outlook}

The Relativistic Quasiparticle Time Blocking Approximation (RQTBA) has been
developed and applied for nuclear structure calculations. The physical content
of this approach is the quasiparticle-vibration coupling model based on the
relativistic energy density functional and the relativistic QRPA. The approach
is formulated for a system with an even particle number in terms of the
Bethe-Salpeter equation for the $ph$-channel in the doubled space to describe
a response of the system in an external field and its spectral characteristics.

The static part of the single-quasiparticle self-energy is determined by the
relativistic energy functional with the parameter set NL3 based on a one-meson
exchange interaction with a non-linear self-coupling between the mesons. An
independent phenomenologically parameterized term is introduced into the
relativistic energy functional to describe pairing correlations which are
considered to be a non-relativistic effect and treated in terms of
Bogoliubov's quasiparticles and, in the application, within the BCS
approximation. In order to take the QPC into account in a consistent way, we
have first calculated the amplitudes of this coupling within the
self-consistent RQRPA with the static interaction. Then, the calculated QPC
energy-dependent self-energy was introduced into the Dirac-Hartree-Bogoliubov
equation for the single-quasiparticle wave function and into the equivalent
Dyson equation for a single-quasiparticle Green's function. The Bethe-Salpeter
equation for the response function in the doubled space contains the
energy-dependent induced interaction connected with the energy-dependent
self-energy by the consistency condition. The BSE has been formulated and
solved in both Dirac-Hartree-BCS and momentum-channel representations.

In order to solve the BSE in the quasiparticle time blocking approximation,
the time-projection technique is used to block the two-quasiparticle
propagation through the states which have more complicated structure than
$2q\otimes phonon$. The nuclear response is then explicitly calculated on the
$2q\otimes phonon$ level by summation of infinite series of Feynman's diagrams.
In order to avoid double counting of the QPC effects a proper subtraction of
the static QPC contribution has been performed. Since the parameters of
density functional for the static RHB description have been adjusted to
experiment they include already essential ground state correlations.

The RQTBA introduced in the first sections of this work is applied
for the calculation of spectroscopic characteristics of the
isovector dipole excitations in the wide energy range up to 30 MeV
for spherical open-shell nuclei, in particular for the isotopes
$^{100,106,114,116,120,130}$Sn (Z=50) and $^{88}$Sr, $^{90}$Zr,
$^{92}$Mo (N=50). The QPC leads to a significant spreading width of
the GDR as compared to RQRPA calculations and causes the strong
fragmentation of the pygmy dipole mode and its spreading to lower
energies. This is in an agreement with experimental data as well as
with the results obtained within the non-relativistic approaches.

The good agreement of our results with the experimental data obtained without
any additional adjustable parameters for a large number of semi-magic nuclei,
confirms the universality of the RMF energy functional and the predictive
power of our approach. We hope, that some of the minor drawbacks in these
calculations can be overcome by using in future an improved version of the
density functional both in the $ph$- and in the $pp$-channel.

\bigskip

\leftline{\bf ACKNOWLEDGEMENTS}

Helpful discussions with Prof. D. Vretenar are gratefully acknowledged. This
work has been supported in part by the Bundesministerium f\"{u}r Bildung und
Forschung under project 06 MT 246 and by the DFG cluster of excellence
\textquotedblleft Origin and Structure of the Universe\textquotedblright%
\ (www.universe-cluster.de). E. L. acknowledges the support from the Alexander
von Humboldt-Stiftung. P.R. thanks for the support provided by the Ministerio
de Educaci\'on y Ciencia, Spain, under the number SAB2005-0025. V.~T.
acknowledges financial support from the Deutsche Forschungsgemeinschaft under
the grant No. 436 RUS 113/806/0-1 and from the Russian Foundation for Basic
Research under the grant No. 05-02-04005-DFG\_a. \bigskip


\end{document}